\documentclass[prb,reprint,twocolumn,amsmath,amssymb]{revtex4-1}

\newcommand{\angstrom}{\text{\normalfont\AA}} 

\newcommand{\bx}{\boldsymbol{\textbf{x}}}

\newcommand{\bR}{\boldsymbol{\textbf{R}}}

\newcommand{\dx}{\,d\bx}

\newcommand{\Rthree}{ {\mathbb{R}^{3}} }

\newcommand{\order}{\mathcal{O}}

\providecommand{\e}[1]{\ensuremath{\times 10^{#1}}}  
\usepackage{graphicx}
\usepackage{graphics}
\usepackage{multirow}
\usepackage{fancyhdr, ifpdf}
\usepackage{amsxtra}
\usepackage{stmaryrd}
\usepackage{mathrsfs}
\usepackage{psfrag}
\usepackage{subfigure}
\usepackage{epsfig}
\usepackage{rotating}
\usepackage{setspace}
\usepackage{float}
\usepackage{amsfonts}
\usepackage{amsmath}
\usepackage{amsbsy}
\usepackage{amscd}
\usepackage{amsthm}
\usepackage{color}
\usepackage{tabularx}
\usepackage{caption}
\usepackage{sidecap}
\usepackage{dcolumn}
\usepackage[ruled]{algorithm2e}
\usepackage{algorithmic}

\definecolor{hellgruen}{rgb}{0.2,0.7,0.2}

\begin{document}
\preprint{}
 \title{Real-space formulation of orbital-free density functional theory using finite-element discretization: The case for Al, Mg, and Al-Mg intermetallics}

\author{Sambit Das}
\author{Mrinal Iyer}
\author{Vikram Gavini}
\affiliation{Department of Mechanical Engineering, University of Michigan, Ann Arbor, MI 48109, USA}


\begin{abstract}
We propose a local real-space formulation for orbital-free DFT with density dependent kinetic energy functionals and a unified variational framework for computing the configurational forces associated with geometry optimization of both internal atomic positions as well as the cell geometry. The proposed real-space formulation, which involves a reformulation of the extended interactions in electrostatic and kinetic energy functionals as local variational problems in auxiliary potential fields, also readily extends to all-electron orbital-free DFT calculations that are employed in warm dense matter calculations. We use the local real-space formulation in conjunction with higher-order finite-element discretization to demonstrate the accuracy of orbital-free DFT and the proposed formalism for the Al-Mg materials system, where we obtain good agreement with Kohn-Sham DFT calculations on a wide range of properties and benchmark calculations. Finally, we investigate the cell-size effects in the electronic structure of point defects, in particular a mono-vacancy in Al. We unambiguously demonstrate that the cell-size effects observed from vacancy formation energies computed using periodic boundary conditions underestimate the extent of the electronic structure perturbations created by the defect. On the contrary, the bulk Dirichlet boundary conditions, accessible only through the proposed real-space formulation, which correspond to an isolated defect embedded in the bulk, show cell-size effects in the defect formation energy that are commensurate with the perturbations in the electronic structure. Our studies suggest that even for a simple defect like a vacancy in Al, we require cell-sizes of $\sim 10^3$ atoms for convergence in the electronic structure. 
\end{abstract}

\maketitle

\section{Introduction}\label{sec:intro}
Electronic structure calculations have played an important role in understanding the properties of a wide range of materials systems~\cite{Martin}. In particular, the Kohn-Sham formalism of density functional theory~\cite{kohn64,kohn65} has been the workhorse of ground-state electronic structure calculations. However, the Kohn-Sham approach requires the computation of single-electron wavefunctions to compute the kinetic energy of non-interacting electrons, whose computational complexity typically scales as $\order(N^3)$ for an $N$-electron system, thus, limiting standard calculations to materials systems containing few hundreds of atoms. While there has been progress in developing close to linear-scaling algorithms for the Kohn-Sham approach~\cite{goedecker99,bow2012}, these are still limited to a few thousands of atoms, especially for metallic systems~\cite{Motamarri2014}. The orbital-free approach to DFT~\cite{ParrYang}, on the other hand, models the kinetic energy of non-interacting electrons as an explicit functional of the electron density, thus circumventing the computationally intensive step of computing the single-electron wavefunctions. Further, the computational complexity of orbital-free DFT scales linearly with the system size as the ground-state DFT problem reduces to a minimization problem in a single field---the electron density. The past two decades has seen considerable progress in the development of accurate models for orbital-free kinetic energy functionals~\cite{Teter1992,Madden1994,Yan1998,Yan1999,Karasiev2012,Ke2013,Karasiev2014,Ke2014}, and, in particular, for systems whose electronic-structure is close to a free electron gas (for e.g. Al, Mg). Also, orbital-free DFT calculations are being increasingly used in the simulations of warm dense matter where the electronic structure is close to that of a free electron gas at very high temperatures~\cite{Flavian2006,Flavian2008,Collins2009,Collins2013,Sheppard2014}. As the reduced computational complexity of orbital-free DFT enables consideration of larger computational domains, recent studies have also focused on studying extended defects in Al and Mg, and have provided important insights into the energetics of these defects~\cite{Gavini2007c,Ho2007,Gang2008,Shin2009,Shin2013,Mrinal2015}.  

The widely used numerical implementation of orbital-free DFT is based on a Fourier space formalism using a plane-wave discretization~\cite{PROFESS2,PROFESS3}. A Fourier space formulation provides an efficient computation of the extended interactions arising in orbital-free DFT---electrostatics and kinetic energy functionals---through Fourier transforms. Further, the plane-wave basis is a complete basis and provides variational convergence in ground-state energy with exponential convergence rates. However, the Fourier space formulations are restricted to periodic geometries and boundary conditions that are suitable for perfect bulk materials, but not for materials systems containing extended defects. Also, the extended spatial nature of the plane-wave basis affects the parallel scalability of the numerical implementation and is also not suitable for multi-scale methods that rely on coarse-graining. In order to address the aforementioned limitations of Fourier space techniques, recent efforts have focussed on developing real-space formulations for orbital-free DFT and numerical implementations based on finite-element~\cite{Gavini2007,Bala2010,Mrinal2012} and finite difference discretizations~\cite{Carlos,Phanish,Phanish2}. 

In the present work, we build on these prior efforts to develop an efficient real-space formulation of orbital-free DFT employing the widely used non-local Wang-Govind-Carter (WGC)~\cite{Yan1999} kinetic energy functional. As in prior efforts~\cite{Gavini2007,Bala2010}, we reformulate the extended interactions in electrostatics and the non-local terms in the WGC kinetic energy functionals as local variational problems in auxiliary potential fields. However, the proposed reformulation of electrostatic interactions is notably different from previous works, and enables the evaluation of variational configurational forces corresponding to both internal atomic relaxations as well as external cell relaxation under a single framework. Further, the proposed formulation naturally extends to all-electron orbital-free DFT calculations of warm dense matter~\cite{Flavian2006,Flavian2008}. In the proposed real-space formulation, the ground-state orbital-free DFT problem is reformulated as an equivalent saddle point problem of a local functional in electron density, electrostatic potential and the auxiliary potential fields (kernel potentials) accounting for the extended interactions in the kinetic energy functional. We employ a higher-order finite-element basis to discretize the formulation, and demonstrate the optimal numerical convergence of both the ground-state energy and configurational forces with respect to the discretization. Further, we propose an efficient numerical approach to compute the saddle point problem in electron density, electrostatic potential and kernel potentials by expressing the saddle point problem as a fixed point iteration problem, and using a self-consistent field approach to solve the fixed point iteration problem.    

We subsequently investigate the accuracy and transferability of the proposed real-space formulation of orbital-free DFT for Al and Mg materials systems. To this end, we compute the bulk properties of Al, Mg and Al-Mg intermetallics, and compare it with Kohn-Sham DFT. As orbital-free DFT only admits local pseudopotentials, the Kohn-Sham DFT calculations are conducted using both local and non-local psedupotentials. Our studies indicates that the bulk properties computed using orbital-free DFT for Al, Mg and Al-Mg intermetallics are in good agreement with Kohn-Sham DFT. We further investigate the accuracy of orbital-free DFT by computing the interatomic forces in Al and Mg, which are also in good agreement with Kohn-Sham DFT calculations. Our studies demonstrate that orbital-free DFT is accurate and transferable across a wide range of properties for Al, Mg and Al-Mg intermetallics, and can be used to study properties of these materials systems that require computational domains that are not accessible using Kohn-Sham DFT. For instance, in the present study we computed the formation energy of $\beta'$ Al-Mg alloy containing $879$ atoms in a unit cell employing the proposed real-space formulation of orbital-free DFT, but the same system was found to be prohibitively expensive using Kohn-Sham DFT.    

We finally investigate the cell-size effects in the electronic structure of point defects, in particular a mono-vacancy in Al. Prior studies using Fourier-based formulations of orbital-free DFT have suggested that the formation energy of a mono-vacancy in Al is well converged by 108-256 atom cell-sizes~\cite{Ho2007}. However, coarse-grained real-space calculations have suggested that much larger cell-sizes of the order of 1,000 atoms are required for convergence of vacancy formation energies~\cite{Bala2010}, which was also supported by asymptotic estimates~\cite{Gavini2011}. In order to understand the underpinnings of this discrepancy, we use the finite-element discretized real-space formulation of orbital-free DFT and compute the vacancy formation energy using two boundary conditions: (i) periodic boundary conditions, equivalent to Fourier-space based formulations; (ii) bulk Dirichlet boundary conditions, where the perturbations in the electronic structure arising due to the vacancy vanishes on the boundary of the computational domain. Our study suggests that while the vacancy formation energy is well converged by 108 atom cell-size using periodic boundary conditions, the electronic fields are not well-converged by this cell-size. On the other hand the bulk Dirichlet boundary conditions show well converged formation energy as well as electronic fields by cell sizes of $\sim$1,000 atoms, which is consistent with prior real-space calculations. This study reveals that while periodic boundary conditions show a superior convergence in formation energies due to the variational nature of the formalism, the true cell-size effects which also measure convergence of electronic fields are provided by the bulk Dirichlet boundary conditions. We note that the proposed real-space formulation with finite-element discretization are crucial to employing bulk Dirichlet boundary conditions, which enable the study of isolated defects in bulk.         

The remainder of the paper is organized as follows. Section II provides a description of the orbital-free DFT problem. Section III presents the proposed real-space formulation of the orbital-free DFT problem, the configurational forces associated with structural relaxations, and the finite-element discretization of the formulation. Section IV discusses the numerical implementation of the formulation and presents an efficient numerical approach for the solution of the saddle point real-space variational problem. Section V presents the numerical convergence results of the finite-element discretization of the real-space formulation, the accuracy and transferability of the real-space orbital-free DFT formalism for Al-Mg materials system, and the study of the role of boundary conditions on the cell-size effects in electronic structure calculations of point defects. We finally conclude with a summary and outlook in Section VI.

\section{Orbital-free density functional theory}\label{sec:OFDFT}
The ground-state energy of a charge neutral materials system containing $M$ nuclei and $N$ valence electrons in density functional theory is given by~\cite{ParrYang,Martin}
\begin{equation}\label{eq:DFT}
E(\rho,\bR)=T_s(\rho)+E_{xc}(\rho)+E_H(\rho)+E_{ext}(\rho,\bR)+E_{zz}(\bR)\,,
\end{equation}
where $\rho$ denotes the electron-density and $\bR=\{\bR_{1},\bR_{2},\ldots,\bR_{M}\}$ denotes the vector containing the positions of $M$ nuclei. In the above, $T_s$ denotes the kinetic energy of non-interacting electrons, $E_{xc}$ is the exchange-correlation energy, $E_{H}$ is the Hartree energy or classical electrostatic interaction energy between electrons, $E_{ext}$ is the classical electrostatic interaction energy between electrons and nuclei, and $E_{zz}$ denotes the electrostatic repulsion energy between nuclei. We now discuss the various contributions to the ground-state energy, beginning with the exchange-correlation energy. 

The exchange-correlation energy, denoted by $E_{xc}$, incorporates all the quantum-mechanical interactions in the ground-state energy of a materials system. While the existence of a universal exchange-correlation energy as a functional of electron-density has been established by Hohenberg, Kohn and Sham~\cite{kohn64,kohn65}, its exact functional form has been elusive to date, and various models have been proposed over the past decades. For solid-state calculations, the local density approximation (LDA)~\cite{alder,perdew} and the generalized gradient approximation~\cite{gga1,gga2} have been widely adopted across a range of materials systems. In particular, the LDA exchange-correlation energy, which is adopted in the present work, has the following functional form:
\begin{equation}\label{eq:exc}
E_{\text{xc}}(\rho) = \int \varepsilon_{\text{xc}}(\rho)\rho(\bx) \dx\,\,,
\end{equation}
where $\varepsilon_{\text{xc}}(\rho)=\varepsilon_x(\rho)+\varepsilon_c(\rho)$, and
\begin{equation}
\varepsilon_x(\rho) = -\frac{3}{4}\left(\frac{3}{\pi}\right)^{1/3}\rho^{1/3}(\bx) \,\,,
\end{equation}
\begin{equation}\label{eq:corr}
\varepsilon_c(\rho) = \begin{cases}
&\frac{\gamma}{(1 + \beta_1\sqrt(r_s) + \beta_2r_s)}\;\;\;\;\;\;\;\;\;\;\;\;\;\;\;\;\;\;\;\;\;\;\;r_s\geq1,\\
&A\,\log r_s + B + C\,r_s\log r_s + D\,r_s\;\;\;\;\;\;\;\;r_s\,<\,1,
\end{cases}
\end{equation}
and $r_s = (3/4\pi\rho)^{1/3}$. In the present work, we use the Ceperley and Alder constants~\cite{perdew} in equation~\eqref{eq:corr}.

The last three terms in equation~\eqref{eq:DFT} represent electrostatic interactions between electrons and nuclei. The Hartree energy, or the electrostatic interaction energy between electrons, is given by
\begin{equation}
E_{H}(\rho) = \frac{1}{2}\int\int\frac{\rho(\bx)\rho(\bx')}{|\bx - \bx'|}\dx\dx'\,.\label{eq:hartree}
\end{equation}
The interaction energy between electrons and nuclei, in the case of local pseudopotentials that are adopted in the present work, is given by
\begin{eqnarray}\label{eq:external}
E_{ext}(\rho,\bR) &=& \int \rho(\bx) V_{ext}(\bx,\bR) \dx \notag\\
&=&\sum_{J}\int \rho(\bx) V^{J}_{ps}(|\bx-\bR_J|) d\bx\,,
\end{eqnarray}
where $V^{J}_{ps}$ denotes the pseudopotential corresponding to the $J^{th}$ nucleus, which, beyond a core radius is the Coulomb potential corresponding to the effective nuclear charge on the $J^{th}$ nucleus. The nuclear repulsive energy is given by
\begin{equation}
E_{zz}(\bR) = \frac{1}{2}\sum_{I}\sum_{J,J \neq I} \frac{Z_I Z_J}{|\bR_I-\bR_J|}\,, \label{eq:repulsive}
\end{equation}
where $Z_I$ denotes the effective nuclear charge on the $I^{th}$ nucleus. The above expression assumes that the core radius of the pseudopotential is smaller than internuclear distances, which is often the case in most solid-state materials systems. We note that in a non-periodic setting, representing a finite atomic system, all the integrals in equations~\eqref{eq:hartree}-\eqref{eq:external} are over $\Rthree$ and the summations in equations~\eqref{eq:external}-\eqref{eq:repulsive} include all the atoms. In the case of an infinite periodic crystal, all the integrals over $\bx$ in equations ~\eqref{eq:hartree}-\eqref{eq:external} are over the unit cell whereas the integrals over $\bx'$ are over $\Rthree$. Similarly, in equations~\eqref{eq:external}-\eqref{eq:repulsive}, the summation over $I$ is on the atoms in the unit cell, and the summation over $J$ extends over all lattice sites. Henceforth, we will adopt these notions for the domain of integration and summation. 

The remainder of the contribution to the ground-state energy is the kinetic energy of non-interacting electrons, denoted by $T_s$, which is computed exactly in the Kohn-Sham formalism by computing the single-electron wavefunctions (eigenfunctions) in the mean-field~\cite{Martin}. The conventional solution of the Kohn-Sham eigenvalue problem, which entails the computation of the lowest $N$ eigenfunctions and eigenvalues of the Kohn-Sham Hamiltonian, scales as $O(N^3)$ that becomes prohibitively expensive for materials systems containing a few thousand atoms. While efforts have been focused towards reducing the computational complexity of the Kohn-Sham eigenvalue problem~\cite{goedecker99,bow2012}, this remains a significant challenge especially in the case of metallic systems. In order to avoid the computational complexity of solving for the wavefunctions to compute $T_s$, the orbital-free approach to DFT models the kinetic energy of non-interacting electrons as an explicit functional of electron density~\cite{ParrYang}. These models are based on theoretically known properties of $T_{s}$ for a uniform electron gas, perturbations of uniform electron gas, and the linear response of uniform electron gas~\cite{ParrYang,Teter1992,Madden1994,Yan1998,Yan1999}. As the orbital-free models for the kinetic energy functional are based on properties of uniform electron gas, their validity is often limited to materials systems whose electronic structure is close to a free electron gas, in particular, the alkali and alkali earth metals. Further, as the orbital-free approach describes the ground-state energy as an explicit functional of electron-density, it limits the pseudopotentials calculations to local pseudopotentials. While these restrictions constrain the applicability of the orbital-free approach, numerical investigations~\cite{Yan1999,Huang2008} indicate that recently developed orbital-free kinetic energy functionals and local pseudopotentials can provide good accuracy for Al and Mg, which comprise of technologically important materials systems. Further, there are ongoing efforts in developing orbital-free kinetic energy models for covalently bonded systems and transition metals~\cite{Xia2012,Huang2012}.

In the present work, we restrict our focus to the Wang-Goving-Carter (WGC) density-dependent orbital-free kinetic energy functional~\cite{Yan1999}, which is a widely used kinetic energy functional for ground-state calculations of materials systems with an electronic structure close to a free electron gas. In particular, the functional form of the WGC orbital-free kinetic energy functional is given by
\begin{equation}\label{eqn:KE}
T_s(\rho)=C_F\int \rho^{5/3}(\bx)\,d\bx + \frac{1}{2}\int |\nabla \sqrt{\rho(\bx)}|^2\,d\bx + T_{K}(\rho)
\end{equation} 
where
\begin{eqnarray}
T_{K}(\rho)=C_F\int\int \rho^{\alpha}(\bx)\,K(\xi_{\gamma}(\bx,\bx'),|\bx-\bx'|)\,\rho^{\beta}(\bx')\,d\bx\,d\bx'\,,\notag\\
\xi_{\gamma}(\bx,\bx')=\Big(\frac{k_F^{\gamma}(\bx)+k_F^{\gamma}(\bx')}{2}\Big)^{1/\gamma}, \quad k_F(\bx)=\big(3\pi^2\rho(\bx)\big)^{1/3}\,.\notag
\end{eqnarray}
In equation~\eqref{eqn:KE}, the first term denotes the Thomas-Fermi energy with $C_F=\frac{3}{10}(3\pi^2)^{2/3}$, and the second term denotes the von-Weizs$\ddot{a}$cker correction~\citep{ParrYang}. The last term denotes the density dependent kernel energy, $T_{K}$, where the kernel $K$ is chosen such that the linear response of a uniform electron gas is given by the Lindhard response~\cite{Finnis}. In the WGC functional~\citep{Yan1999}, the parameters are chosen to be $\{\alpha,\beta\}=\{5/6+\sqrt{5}/6,5/6-\sqrt{5}/6\}$ and $\gamma=2.7$. For materials systems whose electronic structure is close to a free-electron gas, the Taylor expansion of the density dependent kernel about a reference electron density ($\rho_0$), often considered to be the average electron density of the bulk crystal, is employed and is given by
\begin{widetext}
\begin{eqnarray}\label{eq:ker}
\begin{split}
K(\xi_{\gamma}(\bx,\bx'),|\bx-\bx'|) =\,& K_0(|\bx-\bx'|)+K_1(|\bx-\bx'|)\big(\Delta\rho(\bx)+\Delta\rho(\bx')\big)+\frac{1}{2}K_{11}(|\bx-\bx'|)\big((\Delta\rho(\bx))^2+(\Delta\rho(\bx'))^2\big)\\
& + K_{12}(|\bx-\bx'|)\Delta\rho(\bx)\Delta\rho(\bx')+\ldots \,\,.
\end{split}
\end{eqnarray}
\end{widetext}
In the above equation, $\Delta\rho(\bx) = \rho(\bx)-\rho_0$ and the density independent kernels resulting from the Taylor expansion are given by
\begin{eqnarray}
K_0(|\bx-\bx'|) = K(\xi_{\gamma},|\bx-\bx'|)\Big|_{\rho=\rho_0}\notag\\
K_1(|\bx-\bx'|) = \frac{\partial K(\xi_{\gamma},|\bx-\bx'|)}{\partial \rho(\bx)}\Big|_{\rho=\rho_0}\notag\\
K_{11}(|\bx-\bx'|) = \frac{\partial^2 K(\xi_{\gamma},|\bx-\bx'|)}{\partial \rho^2(\bx)}\Big|_{\rho=\rho_0}\notag\\
K_{12}(|\bx-\bx'|) = \frac{\partial^2 K(\xi_{\gamma},|\bx-\bx'|)}{\partial \rho(\bx) \partial \rho(\bx')}\Big|_{\rho=\rho_0}\notag\\
\ldots
\end{eqnarray}  
Numerical investigations have suggested that the Taylor expansion to second order provides a good approximation of the density dependent kernel for materials systems with electronic structure close to a free electron gas~\cite{Choly2002,Yan1999}. In particular, in the second order Taylor expansion, the contribution from $K_{12}$ has been found to dominate contributions from $K_{11}$. Thus, in practical implementations, often, only contributions from $K_{12}$ in the second order terms are retained for computational efficiency.

\section{Real-space formulation of orbital-free DFT}\label{sec:RS-OFDFT}
In this section, we present the local variational real-space reformulation of orbital-free DFT, the configurational forces associated with internal ionic relaxations and cell relaxation, and the finite-element discretization of the formulation. 

\subsection{Local real-space formulation}~\label{sec:RS-formulation}
We recall that the various components of the ground-state energy of a materials system (cf. section~\ref{sec:OFDFT}) are local in real-space, except the electrostatic interaction energy and the kernel energy component of the WGC orbital-free kinetic energy functional that are extended in real-space. Conventionally, these extended interactions are computed in Fourier space to take advantage of the efficient evaluation of convolution integrals using Fourier transforms. For this reason, Fourier space formulations have been the most popular and widely used in orbital-free DFT calculations~\cite{PROFESS2,PROFESS3}. However, Fourier space formulations employing the plane-wave basis result in some significant limitations. Foremost of these is the severe restriction of periodic geometries and boundary conditions. While this is not a limitation in the study of bulk properties of materials, this is a significant limitation in the study of defects in materials. For instance, the geometry of a single isolated dislocation in bulk is not compatible with periodic geometries, and, thus, prior electronic structure studies have mostly been limited to artificial dipole and quadrapole arrangements of dislocations. Further, numerical implementations of Fourier-space formulations also suffer from limited scalability on parallel computing platforms. Moreover, the plane-wave discretization employed in a Fourier space formulation provides a uniform spatial resolution, which is not suitable for the development of coarse-graining techniques---such as the quasi-continuum method~\cite{QCOFDFT}---that rely on an adaptive spatial resolution of the basis. 

We now propose a real-space formulation that is devoid of the aforementioned limitations of a Fourier space formulation. The proposed approach, in spirit, follows along similar lines as recent efforts~\cite{Gavini2007, Bala2010}, but the proposed formulation differs importantly in the way the extended electrostatic interactions are treated. In particular, the proposed formulation provides a unified framework to compute the configurational forces associated with both internal ionic and cell relaxations discussed in~\ref{sec:ConfigurationalForces}. 

We begin by considering the electrostatic interactions that are extended in the real-space. We denote by $\tilde{\delta}(\bx-\bR_I)$ a regularized Dirac distribution located at $\bR_I$, and the $I^{th}$ nuclear charge is given by the charge distribution $-Z_I \tilde{\delta}(\bx-\bR_I)$. Defining $\rho_{nu} (\bx)=-\sum_{I} Z_I \tilde{\delta}(|\bx-\bR_{I}|)$ and $\rho_{nu} (\bx')=-\sum_{J} Z_J \tilde{\delta}(|\bx'-\bR_J|)$, the repulsive energy $E_{zz}$ can subsequently be reformulated as 
\begin{equation}\label{eq:repulsiveEnergy}
E_{zz} = \frac{1}{2}\int\int \frac{\rho_{nu}(\bx)\rho_{nu}(\bx')}{|\bx-\bx'|} d\bx d\bx' - E_{self} \,,
\end{equation}
where $E_{self}$ denotes the self energy of the nuclear charges and is given by
\begin{equation}\label{eq:selfEnergy}
E_{self}=\frac{1}{2}\sum_{I}\int\int \frac{Z_I\tilde{\delta}(|\bx-\bR_{I}|)Z_I\tilde{\delta}(|\bx'-\bR_{I}|)}{|\bx-\bx'|} d\bx d\bx' \,.
\end{equation}
We denote the electrostatic potential corresponding to the $I^{th}$ nuclear charge ($-Z_I\tilde{\delta}(|\bx'-\bR_{I}|)$) as $\bar{V}_{\tilde{\delta}}^{I}(\bx)$, and is given by
\begin{equation}\label{eq:selfPotential}
\bar{V}^{I}_{\tilde{\delta}}(\bx) = -\int \frac{Z_I\tilde{\delta}(|\bx'-\bR_{I}|)}{|\bx-\bx'|} d\bx'\,.
\end{equation} 
The self energy, thus, can be expressed as
\begin{equation}\label{eq:selfEnergy2}
E_{self} = -\frac{1}{2}\sum_{I}\int Z_I\tilde{\delta}(|\bx-\bR_{I}|)\bar{V}^{I}_{\tilde{\delta}}(\bx) d\bx\,.
\end{equation} 
Noting that the kernel corresponding to the extended electrostatic interactions in equations~\eqref{eq:selfEnergy}-\eqref{eq:selfPotential} is the Green's function of the Laplace operator, the electrostatic potential and the electrostatic energy can be computed by taking recourse to the solution of a Poisson equation, or, equivalently, the following local variational problem:
\begin{widetext}
\begin{subequations}\label{eq:selfEnergyLocal}
\begin{equation}
E_{self} = -\sum_{I} \min_{V^{I}\in H^1(\Rthree)} \Big\{\frac{1}{8\pi}\int |\nabla V^{I}(\bx)|^2 d\bx + \int Z_I\tilde{\delta}(|\bx-\bR_{I}|) V^I(\bx) d\bx \Big\}\,,
\end{equation}
\begin{equation}
\bar{V}^{I}_{\tilde{\delta}}(\bx) = arg\,\min_{V^{I}\in H^1(\Rthree)} \Big\{\frac{1}{8\pi}\int |\nabla V^{I}(\bx)|^2 d\bx + \int Z_I\tilde{\delta}(|\bx-\bR_{I}|) V^I(\bx) d\bx \Big\}\,.
\end{equation}
\end{subequations}
\end{widetext}
In the above, $H^1(\Rthree)$ denotes the Hilbert space of functions such that the functions and their first-order derivatives are square integrable on $\Rthree$.

We next consider the electrostatic interaction energy corresponding to both electron and nuclear charge distribution. We denote this by $J(\rho,\rho_{nu})$, which is given by
\begin{equation}
J(\rho,\rho_{nu}) = \frac{1}{2}\int\int \frac{\big(\rho(\bx)+\rho_{nu}(\bx)\big)\big(\rho(\bx')+\rho_{nu}(\bx')\big)}{|\bx-\bx'|} d\bx d\bx'\,.
\end{equation}  
We denote the electrostatic potential corresponding to the total charge distribution (electron and nuclear charge distribution) as $\bar{\phi}$, which is given by
\begin{equation}
\bar{\phi}(\bx) = \int \frac{\rho(\bx')+\rho_{nu}(\bx')}{|\bx-\bx'|}d\bx' \,.
\end{equation}
The electrostatic interaction energy of the total charge distribution, in terms of $\bar{\phi}$, is given by 
\begin{equation}
J(\rho,\rho_{nu}) = \frac{1}{2}\int (\rho(\bx)+\rho_{nu}(\bx))\bar{\phi}(\bx) d \bx \,.
\end{equation}
As before, the electrostatic interaction energy as well as the potential of the total charge distribution can be reformulated as the following local variational problem:
\begin{widetext}
\begin{subequations}\label{eq:TotElecEnergyLocal}
\begin{equation}
J(\rho,\rho_{nu}) = - \min_{\phi\in \mathcal{Y}} \Big\{\frac{1}{8\pi}\int |\nabla \phi(\bx)|^2 d\bx - \int (\rho(\bx)+\rho_{nu}(\bx))\phi(\bx) d\bx \Big\}\,,
\end{equation}
\begin{equation}
\bar{\phi}(\bx) = arg\, \min_{\phi\in \mathcal{Y}} \Big\{\frac{1}{8\pi}\int |\nabla \phi(\bx)|^2 d\bx - \int (\rho(\bx)+\rho_{nu}(\bx))\phi(\bx) d\bx \Big\}\,.
\end{equation}
\end{subequations}
\end{widetext}
In the above, $\mathcal{Y}$ is a suitable function space corresponding to the boundary conditions of the problem. In particular, for non-periodic problems such as isolated cluster of atoms $\mathcal{Y}=H^1(\Rthree)$. For periodic problems, $\mathcal{Y}=H^1_{per}(Q)$ where $Q$ denotes the unit cell and $H^1_{per}(Q)$ denotes the space of periodic functions on $Q$ such that the functions and their first-order derivatives are square integrable.  

The electrostatic interaction energy in DFT, comprising of $E_H$, $E_{ext}$ and $E_{zz}$ (cf. equations~\eqref{eq:hartree}-\eqref{eq:repulsive}), can be rewritten in terms of $J(\rho,\rho_{nu})$ and $E_{self}$ as 
\begin{widetext}
\begin{equation}
E_{H}(\rho)+E_{ext} (\rho, \bR) +E_{zz}(\bR) = J(\rho,\rho_{nu})+\sum_{J}\int (V^{J}_{ps}(|\bx-\bR_{J}|)-\bar{V}^{J}_{\tilde{\delta}}(|\bx-\bR_{J}|))\rho(\bx) d\bx - E_{self} \,.
\end{equation}
 \end{widetext}
For the sake of convenience of representation, we will denote by $\mathcal{V}=\{V^1,V^2,\ldots,V^M\}$ the vector containing the electrostatic potentials corresponding to all nuclear charges in the simulation domain. Using the local reformulation of $J(\rho,\rho_{nu})$ and $E_{self}$ (cf. equations~\eqref{eq:selfEnergyLocal} and \eqref{eq:TotElecEnergyLocal}), the electrostatic interaction energy in DFT can now be expressed as the following local variational problem:
\begin{widetext}
\begin{subequations}\label{eq:ElecRSreformulation}
\begin{equation}
E_{H} + E_{ext} +E_{zz} = \max_{\phi\in \mathcal{Y}} \,\, \min_{V^{I}\in H^1(\Rthree)} \mathcal{L}_{el} (\phi,\mathcal{V},\rho,\mathbf{R})
\end{equation}
\begin{equation}
\begin{split}
\mathcal{L}_{el }(\phi,\mathcal{V},\rho,\mathbf{R}) = & - \frac{1}{8\pi}\int |\nabla \phi(\bx)|^2 d\bx + \int (\rho(\bx)+\rho_{nu}(\bx))\phi(\bx) d\bx + \sum_{J}\int (V^{J}_{ps}(|\bx-\bR_{J}|)-\bar{V}^{J}_{\tilde{\delta}}(|\bx-\bR_{J}|))\rho(\bx) d\bx \\
& + \sum_{I} \left\{\frac{1}{8\pi}\int |\nabla V^{I}(\bx)|^2 d\bx + \int Z_I\tilde{\delta}(|\bx-\bR_{I}|) V^{I}(\bx) d\bx \right\} \,.
\end{split}
\end{equation}
\end{subequations}
 \end{widetext}
In the above, the minimization over $V^{I}$ represents a simultaneous minimization over all electrostatic potentials corresponding to $I=1,2,\ldots,M$. We note that, while the above reformulation of electrostatic interactions has been developed for pseudopotential calculations, this can also be extended to all-electron calculations in a straightforward manner by using $V^{J}_{ps} = \bar{V}^J_{\tilde{\delta}}$ and $Z_I$ to be the total nuclear charge in the above expressions. Thus, this local reformulation provides a unified framework for both pseudopotential as well as all-electron DFT calculations.  

We now consider the local reformulation of the extended interactions in the kernel energy component of the WGC orbital-free kinetic energy functional (cf.~\eqref{eq:ker}). Here we adopt the recently developed local real-space reformulation of the kernel energy~\cite{Bala2010,Mrinal2012}, and recall the key ideas and local reformulation for the sake of completeness. We present the local reformulation of $K_0$ and the local reformulations for other kernels ($K_1$, $K_{11}$, $K_{12}$) follows along similar lines. Consider the kernel energy corresponding to $K_0$ given by
\begin{equation}\label{eq:ker0_energy}
T_{K_0}(\rho)=C_F\int\int \rho^{\alpha}(\bx)\,K_0(|\bx-\bx'|)\,\rho^{\beta}(\bx')\,d\bx\,d\bx'\,.
\end{equation}     
We define potentials $v^0_{\alpha}$ and $v^0_{\beta}$ given by
\begin{eqnarray}\label{eq:kerpotential_v}
v^0_{\alpha}(\bx)=\int K_0(|\bx-\bx'|)\rho^{\alpha}(\bx') d\bx' \,,\notag \\
v^0_{\beta}(\bx)=\int K_0(|\bx-\bx'|)\rho^{\beta}(\bx') d\bx' \,.
\end{eqnarray}
Taking the Fourier transform of the above expressions we obtain
\begin{eqnarray}\label{eq:kerFT}
\widehat{{v}^0_{\alpha}}(\mathbf{k})=\widehat{K_0}(|\mathbf{k}|)\widehat{\rho^{\alpha}}(\mathbf{k})\,,\notag \\
\widehat{{v}^0_{\beta}}(\mathbf{k})=\widehat{K_0}(|\mathbf{k}|)\widehat{\rho^{\beta}}(\mathbf{k})\,.
\end{eqnarray}
Following the ideas developed by Choly \& Kaxiras~\cite{Choly2002}, $\widehat{K_0}$ can be approximated to very good accuracy by using a sum of partial fractions of the following form
\begin{equation}\label{eq:kernelAprrox}
\widehat{K_0}(|\mathbf{k}|)\approx \sum_{j=1}^{m}\frac{A_j|\mathbf{k}|^2}{|\mathbf{k}|^2+B_j},
\end{equation}
where $A_j$, $B_j$, $j=1\ldots m$ are constants, possibly complex, that are determined using a best fit approximation. Using this approximation and taking the inverse Fourier transform of equation~\eqref{eq:kerFT}, the potentials in equation (\ref{eq:kerpotential_v}) reduce to 
\begin{eqnarray}
v^0_{\alpha}(\bx)=\sum\limits_{j=1}^m\,[\omega^0_{\alpha_j}(\bx)+A_j \rho^{\alpha}(\bx)]\,,\notag \\
v^0_{\beta}(\bx)=\sum\limits_{j=1}^m\,[\omega^0_{\beta_j}(\bx )+A_j \rho^{\beta}(\bx)]\,.
\end{eqnarray}
where $\omega^0_{\alpha_j}(\bx)$ and $\omega^0_{\beta_j}(\bx)$ for $j=1\ldots m$ are given by the following Helmholtz equations:
\begin{eqnarray}\label{eq:Helmholtz}
&&-\nabla^2\omega^0_{\alpha_j}+B_j\omega^0_{\alpha_j}+A_jB_j \rho^{\alpha}=0\,,\notag\\
&&-\nabla^2\omega^0_{\beta_j}+B_j\omega^0_{\beta_j}+A_jB_j \rho^{\beta}=0\,.
\end{eqnarray}
We refer to these auxiliary potentials, $\omega^0_{\alpha}=\{\omega^0_{\alpha_1}\ldots \omega^0_{\alpha_m}\}$ and $\omega^0_{\beta}=\{\omega^0_{\beta_1}\ldots \omega^0_{\beta_m}\}$ introduced in the local reformulation of the kernel energy as \emph{kernel potentials}. Expressing the Helmholtz equations in a variational form, we reformulate $T_{K_0}$ in (\ref{eq:ker0_energy}) as the following local variational problem in kernel potentials:
\begin{subequations}\label{eq:kernel_variational}
\begin{equation}
T_{K_0}(\rho)= \min_{\omega^0_{\alpha_j}\in \mathcal{Y}}\max_{\omega^0_{\beta_j}\in \mathcal{Y}}\, \mathcal{L}_{K_0} (\omega^0_{\alpha}, \omega^0_{\beta}, \rho)\,,
\end{equation}
\begin{equation}
\begin{split}
&\mathcal{L}_{K_0} (\omega^0_{\alpha}, \omega^0_{\beta}, \rho) = \sum_{j=1}^{m}C_F\Big\{\int\big[ \frac{1}{A_jB_j}\nabla\omega^0_{\alpha_j}(\bx) \cdot\nabla\omega^0_{\beta_j}(\bx) \\
& + \frac{1}{A_j}\omega^0_{\alpha_j}(\bx)\omega^0_{\beta_j}(\bx) + \omega^0_{\beta_j}(\bx)\rho^{\alpha}(\bx) +\omega^0_{\alpha_j}(\bx)\rho^{\beta}(\bx) \\
& +A_j\rho^{(\alpha+\beta)}(\bx)\big]d\bx\Big\}\,.
\end{split}
\end{equation}
\end{subequations}
The variational problem in equation~\eqref{eq:kernel_variational} represents a simultaneous saddle point problem on kernel potentials $\omega^0_{\alpha_j}$ and $\omega^0_{\beta_j}$ for $j=1,\ldots,m$. Following a similar procedure, we construct the local variational reformulations for the kernel energies $T_{K_1}$, $T_{K_{11}}$ and $T_{K_{12}}$ corresponding to kernels $K_{1}$, $K_{11}$ and $K_{12}$, respectively. We denote by $\mathcal{L}_{K_1}(\omega^1_{\alpha}, \omega^1_{\beta}, \rho)$, $\mathcal{L}_{K_{11}}(\omega^{11}_{\alpha}, \omega^{11}_{\beta}, \rho)$ and $\mathcal{L}_{K_{12}}(\omega^{12}_{\alpha}, \omega^{12}_{\beta}, \rho)$ the Lagrangians with respective kernel potentials corresponding to kernel energies of  $K_{1}$, $K_{11}$ and $K_{12}$, respectively. We refer to the supplemental material for the numerical details of the approximations for each of the kernels used in the present work. 

Finally, using the local variational reformulations of the extended electrostatic and kernel energies, the problem of computing the ground-state energy for a given positions of atoms is given by the following local variational problem in electron-density, electrostatic potentials, and kernel potentials:
\begin{widetext}
\begin{equation}\label{eq:locVar1}
\begin{split}
E_0(\mathbf{R})= \min_{\sqrt{\rho} \in \mathcal{X}} \max_{\phi \in \mathcal{Y}} \min_{\omega^{s}_{\alpha_j}\in \mathcal{Y}}\max_{\omega^{s}_{\beta_j}\in \mathcal{Y}} \, 
\Big\{ & C_F \int {\rho (\bx)^{5/3}}\,d\bx  + \frac{1}{2}\int |\nabla \sqrt{\rho(\bx)}|^2\,d\bx + \int \varepsilon_{\text{xc}}(\rho)\rho(\bx)\dx \\ 
& + \sum_{s}\mathcal{L}_{K_s} (\omega^s_{\alpha}, \omega^s_{\beta}, \rho) + \min_{V^{I}\in H^1(\Rthree)} \mathcal{L}_{el} (\phi,\mathcal{V},\rho,\mathbf{R}) \Big\}\,.
\end{split}
\end{equation}
\end{widetext} 
In the above, $s$ denotes the index corresponding to a kernel, and $\mathcal{X}$ and $\mathcal{Y}$ are suitable function spaces corresponding to the boundary conditions of the problem. In particular, for periodic problems, $\mathcal{Y}=H^1_{per}(Q)$ and $\mathcal{X}=\{\sqrt{\rho}|\sqrt{\rho}\in H^1_{per}(Q), \int \rho=N\}$. It is convenient to use the substitution $u(\bx)=\sqrt{\rho(\bx)}$, and enforce the integral constraint in $\mathcal{X}$ using a Lagrange multiplier. Also, for the sake of notational simplicity, we will denote by $\omega_{\alpha}$ and $\omega_{\beta}$ the array of kernel potentials $\{\omega^0_{\alpha}, \omega^{1}_{\alpha}, \omega^{11}_{\alpha}, \omega^{12}_{\alpha} \}$ and $\{\omega^0_{\beta}, \omega^{1}_{\beta}, \omega^{11}_{\beta}, \omega^{12}_{\beta} \}$, respectively. Subsequently, the variational problem in equation~\eqref{eq:locVar1} can be expressed as 
\begin{widetext}
\begin{eqnarray}\label{eq:locVar2}
E_0(\mathbf{R}) = \min_{u \in \mathcal{Y}} \max_{\phi \in \mathcal{Y}} \min_{\omega^s_{\alpha_j}\in \mathcal{Y}}\max_{\omega^s_{\beta_j}\in \mathcal{Y}}\,\, \mathcal{L} (u,\phi,\omega_{\alpha},\omega_{\beta};\mathbf{R}) \qquad \mbox{subject to}: \int u^2(\bx) \dx = N \,,\\ 
\mathcal{L}(u,\phi,\omega_{\alpha},\omega_{\beta};\mathbf{R}) = \tilde{\mathcal{L}} (u)  + \mathcal{L}_{K} (\omega_{\alpha}, \omega_{\beta}, u^2) + \mathcal{L}_{c} (u,\lambda) +\min_{V^{I}\in H^1(\Rthree)} \mathcal{L}_{el} (\phi,\mathcal{V},u^2,\mathbf{R}) \,,\notag \\
\tilde{\mathcal{L}} (u) = C_F \int {u^{10/3}(\bx)}\,d\bx  + \frac{1}{2}\int |\nabla u(\bx)|^2\,d\bx + \int \varepsilon_{\text{xc}}(u^2) u^2(\bx)\dx \,, \notag \\
\mathcal{L}_{K} (\omega_{\alpha}, \omega_{\beta}, u^2) = \sum_{s} \mathcal{L}_{K_s} (\omega^s_{\alpha}, \omega^s_{\beta}, u^2) \,, \notag \\
\mathcal{L}_{c} (u,\lambda) = \lambda\left(\int u^2(\bx) \dx -N \right) \, .\notag
\end{eqnarray}
\end{widetext}      

\subsection {Configurational forces}~\label{sec:ConfigurationalForces}
We now turn our attention to the configurational forces corresponding to geometry optimization. To this end, we employ the approach of inner variations, where we evaluate the generalized forces corresponding to perturbations of underlying space, which provides a unified expression for the generalized force corresponding to the geometry of the simulation cell---internal atomic positions, as well as, the external cell domain. We consider infinitesimal perturbations of the underlying space $\psi_{\epsilon}: \Rthree \to \Rthree$ corresponding to a generator $\Gamma(\bx)$ given by $\Gamma=\frac{d\psi_{\epsilon}(\bx)}{d\epsilon}|_{\epsilon=0}$ such that $\psi_{0}=I$. We constrain the generator $\Gamma$ such that it only admits rigid body deformations in the compact support of the regularized nuclear charge distribution $\rho_{nu}$ in order to preserve the integral constraint $\int\tilde{\delta}(\bx-\bR_I)d\bx=1$. Let $\bx$ denote a point in $Q$, whose image in $Q'=\psi_{\epsilon}(Q)$ is $\bx'=\psi_{\epsilon}(\bx)$. The ground-state energy on $Q'$ is given by
\begin{eqnarray}
E_{0}(\psi_{\epsilon})= \mathcal{L_{\epsilon}} (u_{\epsilon},\phi_{\epsilon},{\omega_{\alpha}}_{\epsilon},{\omega_{\beta}}_{\epsilon};{\mathbf{R}}_{\epsilon})
\end{eqnarray}
where $u_{\epsilon}$, $\phi_{\epsilon}$, ${\omega_{\alpha}}_{\epsilon}$ and ${\omega_{\beta}}_{\epsilon}$ are solutions of the saddle point variational problem given by equation~\eqref{eq:locVar2} evaluated over the function space $\mathcal{Y'}=H^1_{per}(Q')$. The subscript $\epsilon$ on $\mathcal{L}$ is used to denote that the variational problem is solved on $Q'=\psi_{\epsilon}(Q)$. For the sake of convenience, we will represent the integrand of the Lagrangian $\mathcal{L}$ in equation~\eqref{eq:locVar2} by $f(u,\nabla{u},\phi,\nabla\phi,\omega_{\alpha}, \nabla \omega_{\alpha}, \omega_{\beta}, \nabla\omega_{\beta};V_{ps},\bar{V}_{\tilde{\delta}},\bR)$ and $g(\bar{V}_{\tilde{\delta}}^{I},\nabla \bar{V}_{\tilde{\delta}}^{I};\bR)$, where $f$ denotes the integrand whose integrals are over $Q$ and $g$ denotes the integrand whose integrals are over $\Rthree$. The ground-state energy on $Q'$ in terms of $f$ and $g$ can be expressed as
\begin{eqnarray}
&&E_{0}(\psi_{\epsilon}) = \int_{Q'} f(u_{\epsilon}(\bx'),\nabla_{\bx'}{u}_{\epsilon}(\bx'),\phi_{\epsilon}(\bx'),\nabla_{\bx'}\phi_{\epsilon}(\bx'), {\omega_{\alpha}}_{\epsilon}(\bx'), \notag\\
&& \nabla_{\bx'} {\omega_{\alpha}}_{\epsilon}(\bx'), {\omega_{\beta}}_{\epsilon}(\bx'), \nabla_{\bx'}{\omega_{\beta}}_{\epsilon}(\bx'); V_{ps}(\bx'), \bar{V}_{\tilde{\delta}}(\bx'), \psi_{\epsilon}(\bR)) d\bx'  \notag\\
&&+ \sum_{I}\int_{\Rthree} g(\bar{V}^{I}_{\tilde{\delta}_{\epsilon}}(\bx'),\nabla_{\bx'} \bar{V}^{I}_{\tilde{\delta}_{\epsilon}}(\bx');\psi_{\epsilon}(\bR)) d\bx' \,.
\end{eqnarray}
Transforming the above integral to domain $Q$, we obtain
\begin{eqnarray}
&&E_{0}(\psi_{\epsilon})= \int_{Q}f(u_{\epsilon}(\psi_{\epsilon}(\bx)), \nabla_{\bx}u_{\epsilon}(\psi_{\epsilon}(\bx)).\frac{\partial \bx}{\partial \bx'}, \phi_{\epsilon}(\psi_{\epsilon}(\bx)), \notag\\ &&\nabla_{\bx}\phi_{\epsilon}(\psi_{\epsilon}(\bx)).\frac{\partial \bx}{\partial \bx'}, {\omega_\alpha}_{\epsilon}(\psi_{\epsilon}(\bx)), \nabla_{\bx}{\omega_{\alpha}}_{\epsilon}(\psi_{\epsilon}(\bx)).\frac{\partial \bx}{\partial \bx'}, {\omega_\beta}_{\epsilon}(\psi_{\epsilon}(\bx)), \notag\\
&&  \nabla_{\bx}{\omega_{\beta}}_{\epsilon}(\psi_{\epsilon}(\bx)).\frac{\partial \bx}{\partial \bx'}; V_{ps}(\psi_{\epsilon}(\bx)), \bar{V}_{\tilde{\delta}}(\psi_{\epsilon}(\bx)), \psi_{\epsilon}(\bR)) \det(\frac{\partial \bx'}{\partial \bx})\,d\bx \notag\\
&&+ \sum_{I} \int_{\Rthree} g(\bar{V}^{I}_{\tilde{\delta}_{\epsilon}}(\psi_{\epsilon}(\bx)),\nabla_{\bx} \bar{V}^{I}_{\tilde{\delta}_{\epsilon}}(\psi_{\epsilon}(\bx)).\frac{\partial \bx}{\partial \bx'};\psi_{\epsilon}(\bR)) \det(\frac{\partial \bx'}{\partial \bx})\,d\bx \notag \\   
\end{eqnarray}
We now evaluate the configurational force given by the G\^{a}teaux derivative of $E_{0}(\psi_{\epsilon})$:
\begin{widetext}
\begin{eqnarray}\label{eq:GateuxDerivative}
&&\frac{d E_{0}(\psi_{\epsilon})}{d\epsilon}\Big{|}_{\epsilon=0} = \int_{Q} f(u_{0}(\bx),\nabla{u}_{0}(\bx),\phi_{0}(\bx),\nabla\phi_{0}(\bx), {\omega_{\alpha}}_{0}(\bx),\nabla{\omega_{\alpha}}_{0}(\bx), {\omega_{\beta}}_{0}(\bx), \nabla{\omega_{\beta}}_{0}(\bx);V_{ps}(\bx), \bar{V}_{\tilde{\delta}}(\bx),\bR) \frac{d}{d\epsilon}(\det(\frac{\partial \bx'}{\partial \bx}))\Big{|}_{\epsilon=0} d\bx \notag \\
&&+ \int_{Q}\left(\frac{\partial f}{\partial \nabla u}(\nabla u_{0})\otimes\nabla u_{0} + \frac{\partial f}{\partial \nabla \phi}(\nabla \phi_{0})\otimes\nabla \phi_{0} + \sum_{s} \Big( \frac{\partial f}{\partial \nabla \omega^s_{\alpha}}(\nabla{\omega^s_{\alpha}}_{0})\otimes\nabla {\omega^s_{\alpha}}_{0} + \frac{\partial f}{\partial \nabla \omega^s_{\beta}}(\nabla{\omega^s_{\beta}}_{0})\otimes\nabla {\omega^s_{\beta}}_{0} \Big)\right) : \left(\frac{d}{d \epsilon}\frac{\partial \bx}{\partial \bx'}\Big{|}_{\epsilon=0}\right) d\bx \notag\\
&&+\sum_{J}\int_{Q} u^2_{0}(\bx)\left(\nabla V^{J}_{ps}(|\bx-\bR_{J}|) - \nabla \bar{V}^{J}_{\tilde{\delta}}(|\bx-\bR_{J}|) \right) . \left(\frac{d \psi_{\epsilon}(\bx)}{d\epsilon}\Big{|}_{\epsilon=0}-\frac{d\psi_{\epsilon}(\bR_J)}{d\epsilon}\Big{|}_{\epsilon=0} \right) d\bx \quad \notag \\
&&+ \sum_{I}\int_{\Rthree} g(\bar{V}^{I}_{\tilde{\delta}_{0}}(\bx),\nabla \bar{V}^{I}_{\tilde{\delta}_{0}}(\bx);\bR) \frac{d}{d\epsilon}(\det(\frac{\partial \bx'}{\partial \bx}))\Big{|}_{\epsilon=0} d\bx 
+ \sum_{I}\int_{\Rthree} \frac{\partial g}{\partial \nabla \bar{V}^{I}_{\tilde{\delta}}}(\nabla\bar{V}^{I}_{\tilde{\delta}_{0}})\otimes\nabla \bar{V}^{I}_{\tilde{\delta}_{0}}: \left(\frac{d}{d \epsilon}\frac{\partial \bx}{\partial \bx'}\Big{|}_{\epsilon=0}\right) d\bx \,.
\end{eqnarray}
\end{widetext}
In the above, we denote by `$\otimes$' the outer product between two vector, by `$.$' the dot product between two vectors and by `$:$' the dot product between two tensors. We note that in the above expression there are no terms involving the explicit derivatives of $f$ and $g$ with respect to $\bR$ as $\tilde{\delta}(|\bx'-\psi_{\epsilon}(\bR)|)=\tilde{\delta}(|\bx-\bR|)$, which follows from the restriction that $\psi_{\epsilon}$ corresponds to rigid body deformations in the compact support of $\rho_{nu}$. We further note that terms arising from the inner variations of $E_{0}(\psi_{\epsilon})$ with respect to $u_{\epsilon}$, $\phi_{\epsilon}$, ${\omega_{\alpha}}_{\epsilon}$, ${\omega_{\beta}}_{\epsilon}$ and $\bar{V}^{I}_{\tilde{\delta}_{\epsilon}}$ vanish as $u_{0}$ $\phi_{0}$, ${\omega_{\alpha}}_{0}$, ${\omega_{\beta}}_{0}$ and $\bar{V}^{I}_{\tilde{\delta}_{0}}$ are the solutions of the saddle point variational problem corresponding to $E_{0}(\psi_{0})$. We now note the following identities
\begin{equation}
\begin{split}
\frac{d}{d\epsilon}\left\{\frac{\partial x_i}{\partial x'_j}\right\}\Big{|}_{\epsilon=0}=&-\frac{\partial x_i}{\partial x'_k}\Big(\frac{d}{d\epsilon}\left\{\frac{\partial{\psi_\epsilon}_k}{\partial x_l}\right\}\Big)\frac{\partial x_l}{\partial x'_j}\,\Big{|}_{\epsilon=0}\\
=&-\frac{\partial\Gamma_i}{\partial x_j}\,,
\end{split}
\end{equation}
\begin{equation}
\begin{split}
\frac{d}{d\epsilon}\left\{\det\big(\frac{\partial x'_l}{\partial x_m}\big)\right\}\Big{|}_{\epsilon=0}=& \det\big(\frac{\partial x'_l}{\partial x_m}\big)\frac{\partial x_j}{\partial x'_i}
\Big(\frac{d}{d\epsilon}\left\{\frac{\partial {\psi_\epsilon}_i}{\partial x_j}\right\}\Big)\Big{|}_{\epsilon=0}\\
=&\frac{\partial\Gamma_j}{\partial x_j}.
\end{split}
\end{equation}
Using these identities in equation~\eqref{eq:GateuxDerivative}, and rearranging terms, we arrive at
\begin{eqnarray}\label{eq:Eshelby}
&&\frac{d E_{0}(\psi_{\epsilon})}{d\epsilon}\Big{|}_{\epsilon=0} = \int_{Q}\mathbf{E}:\nabla\Gamma(\bx) \,d\bx + \sum_{I}\int_{\Rthree}{\mathbf{E}^{'}_I}:\nabla\Gamma(\bx) \,d\bx \notag\\
&&+\sum_{J}\int_{Q} u^2_{0}(\bx)\left(\nabla \big(V^{J}_{ps}-\bar{V}^{J}_{\tilde{\delta}}\big) \right).\left( \Gamma(\bx) -\Gamma(\bR_{J})\right) d\bx
\end{eqnarray}
where $\mathbf{E}$ and $\mathbf{E}'$ denote Eshelby tensors corresponding to $f$ and $g$, respectively. The expressions for the Eshelby tensors $\mathbf{E}$ and $\mathbf{E}'_{I}$ explicitly in terms of $u$, $\phi$, $\omega_{\alpha}$, $\omega_{\beta}$, $V_{ps}$ and $\bar{V}_{\tilde{\delta}}$ are given by
\begin{widetext}
\begin{eqnarray}
\mathbf{E} = &&\left(C_Fu^{10/3}+\frac{1}{2}|\nabla{u}|^2+\varepsilon_{xc}(u^2)u^2+\lambda u^2 -\frac{1}{8\pi}|\nabla\phi|^2+u^2\phi +\sum_{J}\big(V^{J}_{ps}-\bar{V}^{J}_{\tilde{\delta}}\big)u^2+\sum_{s}f_{K_s}(\omega_{\alpha}^{s}, \nabla\omega_{\alpha}^{s},\omega_{\beta}^{s}, \nabla\omega_{\beta}^{s}, u^2)\right)\mathbf{I}\notag\\
&&-\nabla{u}\otimes\nabla{u} + \frac{1}{4\pi}\nabla{\phi}\otimes\nabla{\phi} - \sum_{s} \left(\frac{\partial f_{K_s}}{\partial\nabla\omega_{\alpha}^{s}}\otimes\nabla\omega_{\alpha}^{s} + \frac{\partial f_{K_s}}{\partial\nabla\omega_{\beta}^{s}}\otimes\nabla\omega_{\beta}^{s}\right)\\
\mathbf{E}^{'}_{I} = && \frac{1}{8\pi} |\nabla\bar{V}^{I}_{\tilde{\delta}}|^2\mathbf{I}-\frac{1}{4\pi}\nabla\bar{V}^{I}_{\tilde{\delta}}\otimes\nabla\bar{V}^{I}_{\tilde{\delta}}
\end{eqnarray}
\end{widetext}
In the above, for the sake of brevity, we represented by $f_{K_s}$ the integrand corresponding to $\mathcal{L}_{K_s}$. We also note that the terms $\phi\rho_{nu}$ and ${V}^{I}_{\tilde{\delta}}\tilde{\delta}(\bx-\bR_{I})$ do not appear in the expressions for $\mathbf{E}$ and $\mathbf{E}^{'}_{I}$, respectively, as $\nabla.\Gamma=0$ on the compact support of $\rho_{nu}$ owing to the restriction that $\Gamma$ corresponds to rigid body deformations in these regions. It may appear that evaluation of the second term in equation~\eqref{eq:Eshelby} is not tractable as it involves an integral over $\Rthree$. To this end, we split this integral on a bounded domain $\Omega$ containing the compact support of $\tilde{\delta}(\bx-\bR_I)$, and its complement. The integral on $\Rthree/\Omega$ can be computed as a surface integral. Thus,
\begin{eqnarray}
&&\int_{\Rthree}\mathbf{E}^{'}_{I}:\nabla\Gamma \,d\bx=\int_{\Omega}\mathbf{E}^{'}_{I}:\nabla\Gamma \,d\bx+ \int_{\Rthree/\Omega}\mathbf{E}^{'}_{I}:\nabla\Gamma \,d\bx\notag\\
&&= \int_{\Omega}\mathbf{E}^{'}_{I}:\nabla\Gamma \,d\bx - \int_{\partial\Omega} \mathbf{E}^{'}_{I}:\hat{\mathbf{n}}\otimes\Gamma \,d\mathbf{s}\,,
\end{eqnarray}
where $\hat{\mathbf{n}}$ denotes the outward normal to the surface $\partial \Omega$. The last equality follows from the fact that $\nabla^2\bar{V}^{I}_{\tilde{\delta}}=0$ on $\Rthree/\Omega$.

The configurational force in equation~\eqref{eq:Eshelby} provides the generalized variational force with respect to both the internal positions of atoms as well as the external cell domain. In order to compute the force on any given atom, we restrict the compact support of $\Gamma$ to only include the atom of interest. In order to compute the stresses associated with cell relaxation (keeping the fractional coordinates of atoms fixed), we restrict $\Gamma$ to affine deformations. Thus, this provides a unified expression for geometry optimization corresponding to both internal ionic relaxations as well as cell relaxation. We further note that, while we derived the configurational force for the case of pseudopotential calculations, the derived expression is equally applicable for all-electron calculations by using $V^{J}_{ps}=\bar{V}^{J}_{\tilde{\delta}}$.    

\subsection{Finite-element discretization}~\label{sec:FE-discretization}
Among numerical discretization techniques, the plane-wave discretization has been the most popular and widely used in orbital-free DFT~\cite{PROFESS2,PROFESS3} as it naturally lends itself to the evaluation of the extended interactions in electrostatic energy and kernel kinetic energy functionals using Fourier transforms. Further, the plane wave basis offers systematic convergence with exponential convergence in the number of basis functions. However, as noted previously, the plane-wave basis also suffers from notable drawbacks. Importantly, plane-wave discretization is restricted to periodic geometries and boundary conditions which introduces a significant limitation, especially in the study of defects in bulk materials~\cite{Mrinal2015}. Further, the plane-wave basis has a uniform spatial resolution, and thus is not amenable to adaptive coarse-graining. Moreover, the use of plane-wave discretization involves the numerical evaluation of Fourier transforms whose scalability is limited on parallel computing platforms. 

In order to circumvent these limitations of the plane-wave basis, there is an increasing focus on developing real-space discretization techniques for orbital-free DFT based on finite-difference discretization~\cite{Carlos, Phanish, Phanish2} and finite-element discretization~\cite{Gavini2007,Mrinal2012}. In particular, the finite-element basis~\cite{Brenner-Scott}, which is a piecewise continuous polynomial basis,  has many features of a desirable basis in electronic structure calculations. While being a complete basis, the finite-element basis naturally allows for the consideration of complex geometries and boundary conditions, is amenable to unstructured coarse-graining, and exhibits good scalability on massively parallel computing platforms. Moreover, the adaptive nature of the finite-element discretization also enables the consideration of all-electron orbital-free DFT calculations that are widely used in studies of warm dense matter~\cite{Flavian2006,Flavian2008,Collins2013}. Further, recent numerical studies have shown that by using a higher-order finite-element discretization significant computational savings can be realized for both orbital-free DFT~\cite{Mrinal2012} and Kohn-Sham DFT calculations~\cite{Motamarri2013,Motamarri2014}, effectively overcoming the degree of freedom disadvantage of the finite-element basis in comparison to the plane-wave basis. 

Let $\mathcal{Y}_h$ denote the finite-element subspace of $\mathcal{Y}$, where $h$ represents the finite-element mesh size. The discrete problem of computing the ground-state energy for a given positions of atoms, corresponding to equation~\eqref{eq:locVar2}, is given by the constrained variational problem:
\begin{eqnarray}\label{eq:VarFE}
E_0(\mathbf{R}) = &&\min_{u_h \in \mathcal{Y}_h} \max_{\phi_h \in \mathcal{Y}_h} \min_{{{\omega^s_{\alpha_j}}_h}\in \mathcal{Y}_h}\max_{{{\omega^s_{\beta_j}}_h}\in \mathcal{Y}_h}\,\, \mathcal{L} (u_h,\phi_h,{\omega_{\alpha_h}},{\omega_{\beta_h}};\mathbf{R}) \notag \\ 
&&\mbox{subject to}: \int u_h^2(\bx) \dx = N\,.
\end{eqnarray}   
In the above, $u_h$, $\phi_h$, ${\omega_{\alpha}}_h$ and ${\omega_{\beta}}_h$ denote the finite-element discretized fields corresponding to square-root electron-density, electrostatic potential, and kernel potentials, respectively. We restrict our finite-element discretization such that atoms are located on the nodes of the finite-element mesh. In order to compute the finite-element discretized solution of $\bar{V}^{J}_{\tilde{\delta}}$, we represent $\tilde{\delta}(\bx-\bR_J)$ as a point charge on the finite-element node located at $\bR_{J}$, and the finite-element discretization provides a regularization for $\bar{V}^{J}_{\tilde{\delta}}$. Previous investigations have suggested that such an approach provides optimal rates of convergence of the ground-state energy (cf.~\cite{Mrinal2012,Motamarri2013} for a discussion). 

The finite-element basis functions also provide the generator of the deformations of the underlying space in the isoparametric formulation, where the same finite-element shape functions are used to discretize both the spatial domain as well as the fields prescribed over the domain. Thus, the configurational force associated with the location of any node in the finite-element mesh can be computed by substituting for $\Gamma$, in equation~\eqref{eq:Eshelby}, the finite-element shape function associated with the node. Thus, the configurational force on any finite-element node located at an atom location corresponds to the variational ionic force, which are used to drive the internal atomic relaxation. The forces on the finite-element nodes that do not correspond to an atom location represent the generalized force of the energy with respect to the location of the finite-element nodes, and these can be used to obtain the optimal location of the finite-element nodes---a basis adaptation technique.   

We note that the local real-space variational formulation in section~\ref{sec:RS-formulation}, where the extended interactions in the electrostatic energy and kernel functionals are reformulated as local variational problems, is essential for the finite-element discretization of the formulation.

\section{Numerical Implementation}\label{sec:Numerics}
In this section, we present the details of the numerical implementation of the finite-element discretization of the real-space formulation of orbital-free DFT discussed in section~\ref{sec:RS-OFDFT}. Subsequently, we discuss the solution procedure for the resulting discrete coupled equations in square-root electron-density, electrostatic potential and kernel potentials. 

\subsection{Finite-element basis}~\label{sec:FE-basis}
A finite-element discretization using linear tetrahedral finite-elements has been the most widely used discretization technique for a wide range of partial differential equations. Linear tetrahedral elements are well suited for problems involving complex geometries and moderate levels of accuracy. However in electronic structure calculations, where the desired accuracy is commensurate with chemical accuracy, linear finite elements are computationally inefficient requiring of the order of hundred thousand basis functions per atom to achieve chemical accuracy. A recent study~\cite{Mrinal2012} has demonstrated the significant computational savings---of the order of 1000-fold compared to linear finite-elements---that can be realized by using higher-order finite-element discretizations. Thus, in the present work we use higher-order hexahedral finite elements, where the basis functions are constructed as a tensor product of basis functions in one-dimension~\cite{Brenner-Scott}. 

\subsection{Solution procedure}~\label{sec:NumerSoln}
The discrete variational problem in equation~\eqref{eq:VarFE} involves the computation of the following fields---square-root electron-density, electrostatic potential and kernel potentials. Two solution procedures, suggested in prior efforts~\cite{Mrinal2012}, for solving this discrete variational problem include: (i) a simultaneous solution of all the discrete fields in the problem; (ii) a nested solution procedure, where for every trial square-root electron-density the discrete electrostatic and kernel potential fields are computed. Given the non-linear nature of the problem, the simultaneous approach is very sensitive to the starting guess and often suffers from lack of robust convergence, especially for large-scale problems. The nested solution approach, on the other hand, while constituting a robust solution procedure, is computationally inefficient due to the huge computational costs incurred in computing the kernel potentials which involves the solution of a series of Helmholtz equations (cf. equation~\eqref{eq:Helmholtz}). Thus, in the present work, we will recast the local variational problem in equation~\eqref{eq:VarFE} as the following fixed point iteration problem:
\begin{subequations}\label{eq:fixedPoint}
\begin{eqnarray}\label{eq:fixedPoint_a}
\{\bar{u}_h,\bar{\phi}_{h}\} = && \,\,arg\, \min_{u_h}\, arg\, \max_{\phi_h} \mathcal{L}(u_h,\phi_h,\bar{\omega}_{{\alpha}_h}, \bar{\omega}_{{\beta}_h};\bR) \notag\\ 
&&\mbox{subject to}: \int u_h^2(\bx) \dx = N. 
\end{eqnarray}
\begin{equation}\label{eq:fixedPoint_b}
\{\bar{\omega}_{{\alpha}_h}, \bar{\omega}_{{\beta}_h}\} = \,\, arg\, \min_{\omega_{{\alpha}_h}}\, arg\, \max_{\omega_{{\beta}_h}} \mathcal{L}(\bar{u}_h,\bar{\phi}_h,\omega_{{\alpha}_h}, \omega_{{\beta}_h};\bR) \,.
\end{equation}
\end{subequations} 
We solve this fixed point iteration problem using a mixing scheme, and, in particular, we employ the Anderson mixing scheme~\cite{Anderson} with full history in this work. Our numerical investigations suggest that the fixed point iteration converges, typically, in less than ten self-consistent iterations even for large-scale problems, thus, providing a numerically efficient and robust solution procedure for the solution of the local variational orbital-free DFT problem. We note that this idea of fixed point iteration has independently and simultaneously been investigated by another group in the context of finite difference discretization~\cite{Phanish2}, and have resulted in similar findings.      

In the fixed point iteration problem, we employ a simultaneous solution procedure to solve the non-linear saddle point variational problem in $u_h$ and $\phi_h$ (equation~\eqref{eq:fixedPoint_a}). We employ an inexact Newton solver provided by the PETSc package~\cite{PETSC} with field split preconditioning and generalized-minimal residual method (GMRES)~\cite{GMRES} as the linear solver. The discrete Helmholtz equations in equation~\eqref{eq:fixedPoint_b} are solved by employing block Jacobi preconditioning and using GMRES as the linear solver. An efficient and scalable parallel implementation of the solution procedure has been developed to take advantage of the parallel computing resources for conducting the large-scale simulations reported in this work.

\section{Results and Discussion}\label{sec:Results}
In this section, we discuss the numerical studies on Al, Mg and Al-Mg intermetallics to investigate the accuracy and transferability of the real-space formulation of orbital-free DFT (RS-OFDFT) proposed in section~\ref{sec:RS-OFDFT}. Wherever applicable, we benchmark the real-space orbital-free DFT calculations with plane-wave based orbital-free DFT calculations conducted using PROFESS~\cite{PROFESS2}, and compare with Kohn-Sham DFT (KS-DFT) calculations conducted using the plane-wave based ABINIT code~\cite{abinit1,abinit2}. Further, we demonstrate the usefulness of the proposed real-space formulation in studying the electronic structure of isolated defects.

\subsection{General calculation details}~\label{sec:calc}

\begin{figure}[htbp]
\includegraphics[width=0.46\textwidth]{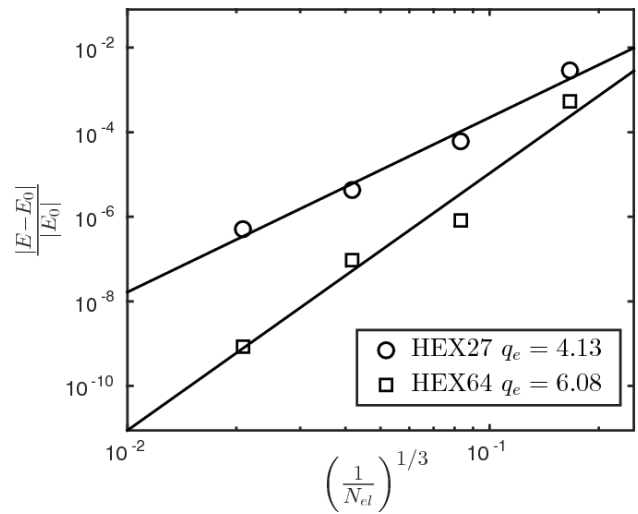}
\caption{\label{fig:energyConv}\small{Convergence of the finite-element approximation in the energy of a fcc Al unit cell with lattice constant $a=7.2$ Bohr.}}
\end{figure}
\begin{figure}[htbp]
\includegraphics[width=0.46\textwidth]{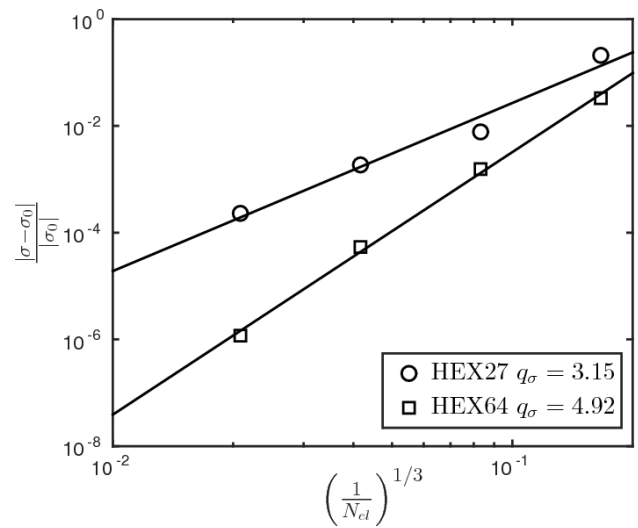}
\caption{\label{fig:stressConv}\small{ Convergence of the finite-element approximation in the hydrostatic stress of a fcc Al unit cell with lattice constant $a=7.2$ Bohr.}}
\end{figure}

In all the real-space orbital-free DFT calculations reported in this section, we use the local reformulation of the density-dependent WGC~\cite{Yan1999} kinetic energy functional proposed in section~\ref{sec:RS-formulation}, the local density approximation (LDA)~\cite{perdew} for the exchange-correlation energy, and bulk derived local pseudopotentials (BLPS)~\cite{Huang2008} for Al and Mg. Cell stresses and ionic forces are calculated using the unified variational formulation of configurational forces developed in section~\ref{sec:ConfigurationalForces}.  In the second order Taylor expansion of the density-dependent WGC functional about the bulk electron density (cf. Section~\ref{sec:OFDFT}), we only retain the $K_{12}$ term for the computation of bulk properties as the contributions from $K_{12}$ dominate those of $K_{11}$ for bulk materials systems. However, in the calculations involving mono-vacancies, where significant spatial perturbations in the electronic structure are present, we use the full second order Taylor expansion of the density dependent WGC functional.  We recall from section~\ref{sec:RS-formulation} that in order to obtain a local real-space reformulation of the extended interactions in the kinetic energy functionals, the kernels ($K_{0}$, $K_{1}$, $K_{11}$, $K_{12}$) are approximated using a sum of $m$ partial fractions where the coefficients of the partial fractions are computed using a best fit approximation (cf. equation~\eqref{eq:kernelAprrox}). These best fit approximations for $m=4,5,6$ that are employed in the present work are given in the supplemental material. It has been shown in recent studies that $m=4$ suffices for Al~\cite{Bala2010,Phanish2}. However, we find that $m=6$ is required to obtain the desired accuracy in the bulk properties of Mg, and Table~\ref{tab:bulkTrf2} shows the comparison between the kernel approximation with $m=6$ and plane-wave based orbital-free DFT calculations conducted using PROFESS~\cite{PROFESS2} for Mg. Thus, we use the best fit approximation of the kernels with $m=4$ for Al, and employ the approximation with $m=6$ for Mg and Al-Mg intermetallics. Henceforth, we will refer by RS-OFDFT-FE the real-space orbital-DFT calculations conducted by employing the local formulation and finite element discretization proposed in section~\ref{sec:RS-OFDFT}. 

The KS-DFT calculations used to assess the accuracy and transferability of the proposed real-space orbital-free DFT formalism are performed using the LDA exchange correlation functional~\cite{perdew}. The KS-DFT calculations are conducted using both local BLPS as well as the non-local Troullier-Martins pseudopotential (TM-NLPS) ~\cite{NLPS} in order to assess the accuracy and transferability of both the model kinetic energy functionals in orbital-free DFT as well as the local pseudopotentials to which the orbital-free DFT formalism is restricted to. The TM-NLPS for Al and Mg are generated using the fhi98PP code ~\cite{FHI98PP}. Within the fhi98PP code, we use the following inputs: $3d$ angular momentum channel as the local pseudopotential component for both Al and Mg, default core cutoff radii for the $3s$, $3p$, and $3d$ angular momentum channels, which are 
$\left \{ 1.790,\, 1.974,\, 2.124 \right \}$ Bohr and $\left\{2.087,\, 2.476,\, 2.476\right\}$ Bohr for Al and Mg respectively, and the  LDA ~\cite{perdew} exchange-correlation. For brevity, henceforth, we refer to the KS-DFT calculations with BLPS and TM-NLPS as KS-BLPS and KS-NLPS, respectively. 

In all the RS-OFDFT-FE calculations reported in this work, the finite-element discretization, order of the finite-elements, numerical quadrature rules and stopping tolerances are chosen such that we obtain 1 meV/ atom accuracy in energies, $1\e{-7} \, \rm{Hartree}\,\, \rm{Bohr}^{-3}$ accuracy in cell stresses and $1\e{-5} \, \rm{Hartree}\,\, \rm{Bohr}^{-1}$ accuracy in ionic forces. Similar accuracies in energies, stresses and ionic forces are achieved for KS-DFT calculations by choosing the appropriate k-point mesh, plane-wave energy cutoff, and stopping tolerances within ABINIT's framework. All calculations involving geometry optimization are conducted until cell stresses and ionic forces are below threshold values of  $5\e{-7} \, \rm{Hartree}\,\, \rm{Bohr}^{-3}$ and $5\e{-5} \, \rm{Hartree}\,\, \rm{Bohr}^{-1}$, respectively.

\subsection{Convergence of finite-element discretization}~\label{sec:Convergence}
We now study the convergence of energy and stresses with respect to the finite-element discretization of the proposed real-space orbital-free DFT formulation. In a prior study on the computational efficiency afforded by higher-order finite-element discretization in orbital-free DFT~\cite{Mrinal2012}, it was shown that second and third-order finite-elements offer an optimal choice between accuracy and computational efficiency. Thus, in the present study, we limit our convergence studies to HEX27 and HEX64 finite-elements, which correspond to second- and third-order finite-elements. As a benchmark system, we consider a stressed fcc Al unit cell with a lattice constant $a=7.2$ Bohr. We first construct a coarse finite-element mesh and subsequently perform a uniform subdivision to obtain a sequence of increasingly refined meshes. We denote by $h$ the measure of the size of the finite-element. For these sequence of meshes, we hold the cell geometry fixed and compute the discrete ground-state energy, $E_{h}$, and hydrostatic stress, $\sigma_{h}$.  The extrapolation procedure proposed in Motamarri et. al~\cite{Mrinal2012} allows us to estimate the ground-state energy and hydrostatic stress in the limit as $h\to 0$, denoted by $E_0$ and $\sigma_0$. To this end, the energy and hydrostatic stress computed from the sequence of meshes using HEX64 finite-elements are fitted to expressions of the form  
\begin{eqnarray}
\left|E_{0}-E_h \right| = C_{e} \left(\frac{1}{N_{el}}\right)^{\frac{q_e}{3}} \,, \notag\\
\left|\sigma_{0}-\sigma_h \right| = C_{\sigma} \left(\frac{1}{N_{el}}\right)^{\frac{q_{\sigma}}{3}} \,, 
\end{eqnarray}
to determine $E_{0},\, q_e,\, \sigma_{0}, \& \, q_{\sigma}$. In the above expression, $N_{el}$ denotes the number of elements in a finite-element mesh. We subsequently use $E_0$ and $\sigma_0$ as the exact values of the ground-state energy and hydrostatic stress, respectively, for the benchmark system. Figures~\ref{fig:energyConv} and~\ref{fig:stressConv} show the relative errors in energy and hydrostatic stress plotted against $\left(\frac{1}{N_{el}}\right)^{\frac{1}{3}}$, which represents a measure of $h$. We note that the slopes of these curves provide the rates of convergence of the finite-element approximation for energy and stresses. These results show that we obtain close to optimal rates of convergence in energy of $\order(h^{2k})$, where $k$ is polynomial interpolation order ($k=2$ for HEX27 and $k=3$ for HEX64). Further, we obtain close to $\order(h^{2k-1})$ convergence in the stresses, which represents optimal convergence for stresses. The results also suggest that higher accuracies in energy and stress are obtained with HEX64 in comparison to HEX27. Thus, we will employ HEX64 finite-elements for the remainder of our study.   
\begin{table}[htbp]
\caption{\label{tab:bulkTrf1} \small{The energy difference in eV between a stable phase and the most stable phase for Al and Mg computed using RS-OFDFT-FE and KS-DFT with TM-NLPS.}}
\begin{ruledtabular}
\begin{tabular}{cccccc}
{Al}  & fcc & hcp & bcc & sc & dia  \\
\hline
RS-OFDFT-FE &  0\footnotemark[1] & 0.016 & 0.075 & 0.339 & 0.843  \\
KS-NLPS & 0  &  0.038 & 0.106  & 0.400  & 0.819   \\
\hline
{Mg}  & hcp & fcc & bcc & sc & dia  \\
\hline
RS-OFDFT-FE &  0 & 0.003 & 0.019 & 0.343 & 0.847  \\
KS-NLPS & 0  &  0.014 & 0.030  & 0.400  & 0.822   \\  
\end{tabular}
\end{ruledtabular}
\footnotetext[1]{The zero in the first column is to indicate that these numbers are the reference against which energies of other phases are determined.}
\end{table}

\begin{table*}[!htpb]
\caption{\label{tab:bulkTrf2} \small{Bulk properties of Al and Mg: Equilibrium ground-state energy per atom ($E_{\rm min}$ in eV), volume per atom ($V_{0}$ in $\angstrom^3$) and bulk modulus ($B_0$ in GPa) computed using RS-OFDFT-FE, PROFESS, and KS-DFT with BLPS and TM-NLPS.}}
\begin{ruledtabular}
\begin{tabular}{ccccc}
Al\footnotemark[1] & RS-OFDFT-FE & PROFESS & KS-BLPS & KS-NLPS\\
\hline
$E_{\rm min}$ & -57.935 & -57.936  &-57.954 & -57.207\\
$V_{0}$ & 15.68 & 15.68 & 15.62 & 15.55\\
$B_{0}$ & 81.7 & 81.5 & 84.1 & 83.6\\
\hline
Mg\footnotemark[2]  & RS-OFDFT-FE & PROFESS & KS-BLPS & KS-NLPS\\
\hline
$E_{\rm min}$ & -24.647 & -24.647 & -24.678 & -24.514\\
$V_{0}$ & 21.40 & 21.43 & 21.18 & 21.26\\
$B_{0}$ & 36.8 & 36.6 & 38.5 & 38.6\\
\end{tabular}
\end{ruledtabular}
\footnotetext[1]{Cell-relaxed lattice constant for fcc Al using RS-OFDFT-FE is $a_0=\, 7.51$ Bohr.}
\footnotetext[2]{Cell-relaxed lattice constants for hcp Mg using RS-OFDFT-FE are $a_0=\, 5.89$ Bohr,\, $c_{0}=\, 9.62$ Bohr.}
\end{table*}

\subsection{Bulk properties of Al, Mg and Al-Mg intermetallics}~\label{sec:BulkTransferability}
We now study the accuracy and transferability of the proposed real-space formulation of orbital-free DFT for bulk properties of Al, Mg and Al-Mg intermetallics. To this end, we begin with the phase stability study of Al and Mg, where we compute the difference in the ground-state energy of a stable phase and the ground-state energy of the most stable phase. The results for Al and Mg are shown in Table~\ref{tab:bulkTrf1}, and are compared against those obtained with KS-DFT employing TM-NLPS. We note that RS-OFDFT-FE correctly predicts the most stable phases of Al and Mg being fcc and hcp, respectively. Further, the stability ordering of the various phases computed using RS-OFDFT-FE is consistent with KS-DFT TM-NLPS calculations. Moreover, the energy differences between the various stable phases and the most stable phase computed using RS-OFDFT-FE are in close agreement with KS-DFT calculations.  

We next consider bulk properties of Al, Mg and Al-Mg intermetallics. To this end, for each system, we first optimize cell geometry and ionic positions to determine the equilibrium cell structure, equilibrium volume ($V_{0}$) and ground-state energy ($E_{\rm min}$). We subsequently compute the bulk modulus given by~\cite{Finnis}
\begin{equation}\label{eq:bulkModulus}
  B= \left.V \frac{\partial^{2} E}{\partial V^2}\right|_{V=V_{0}}\, ,
\end{equation}
where $E$ denotes the ground-state energy of a unit-cell with volume $V$. To compute the bulk modulus, we vary the cell volume by applying a volumetric deformation to the relaxed (equilibrium) unit-cell, which transforms the equilibrium cell vectors $\left\{ {\bf c}_1 \, ,{\bf c}_2 \, , {\bf c}_{3}\right\}$ to $\left\{ {\bf c}^{\prime}_1 \, ,{\bf c}^{\prime}_2 \, , {\bf c}^{\prime}_{3}\right\}$ and are given by 
\begin{equation}
c^{\prime}_{ij}=c_{ij} \, (1+ \eta) \,.
\end{equation} 
While keeping the cell structure fixed, we calculate the ground-state energy for each $\eta$ between $-0.01$ to $0.01$ in steps of 0.002 and fit a cubic polynomial to the $E-V$ data. We subsequently compute the bulk modulus, using equation~\eqref{eq:bulkModulus}, at the equilibrium volume, $V_0$. The computed bulk properties---ground-state energy, equilibrium volume and bulk modulus at equilibrium---for Al and Mg are given in Table~\ref{tab:bulkTrf2}, and those of Al-Mg intermetallics (${\rm Al}_{3}\rm{Mg}$, ${\rm Mg}_{13}{\rm Al}_{14}$, ${\rm Mg}_{17}{\rm Al}_{12}$, and ${\rm Mg}_{23}{\rm Al}_{30}$) are given in Table~\ref{tab:bulkTrf3}. These results suggest that the bulk properties of Al, Mg and Al-Mg intermetallics computed using RS-OFDFT-FE are in good agreement with PROFESS and KS-DFT calculations. 

\begin{table*}[htbp]
\caption{\label{tab:bulkTrf3}\small{Bulk properties of Al-Mg intermetallics: Equilibrium ground-state energy per primitive cell ($E_{\rm min}$ in eV), volume of primitive cell ($V_{0}$ in $\angstrom^3$), and bulk modulus ($B_0$ in GPa) computed using RS-OFDFT-FE, PROFESS, and KS-DFT with BLPS and TM-NLPS.}}
\begin{ruledtabular}
\begin{tabular}{ccccc}
 ${\rm Al}_{3}\rm{Mg}$ & RS-OFDFT-FE & PROFESS &KS-BLPS & KS-NLPS\\
\hline
$E_{\rm min}$ & -198.492 & -198.496& -198.575 & -196.162\\
$V_{0}$ & 67.23 & 67.31&67.13 & 66.52\\
$B_{0}$ & 69.2 &  67.0& 67.6 & 71.0\\
\hline
 ${\rm Mg}_{13}{\rm Al}_{14}$ & RS-OFDFT-FE & PROFESS & KS-BLPS & KS-NLPS\\
\hline
$E_{\rm min}$ & -1130.083 & -1130.100& -1130.972 & -1117.936\\
$V_{0}$ & 494.77 & 494.73& 498.19 & 492.73\\
$B_{0}$ & 53.1 &52.1 & 54.7 & 54.8\\
\hline
 ${\rm Mg}_{17}{\rm Al}_{12}$  & RS-OFDFT-FE & PROFESS & KS-BLPS & KS-NLPS\\
\hline
$E_{\rm min}$ & -1114.446 & -1114.526& -1116.185 & -1104.012\\
$V_{0}$ & 545.32 & 544.85  & 543.67 & 544.21\\
$B_{0}$ & 51.1 & 52.3 & 55.2 & 54.4\\
\hline
${\rm Mg}_{23}{\rm Al}_{30}$ & RS-OFDFT-FE & PROFESS & KS-BLPS & KS-NLPS\\
\hline
$E_{\rm min}$ & -2306.785 & -2306.762 &-2307.989 & -2281.082\\
$V_{0}$ & 953.87 &  952.55& 963.72 & 957.46\\
$B_{0}$ & 64.2 & 60.9 & 60.5 & 60.5\\
\end{tabular}
\end{ruledtabular}
\end{table*}

Finally, we consider the formation energies of Al-Mg intermetallics. In addition to the Al-Mg intermetallics for which we computed the bulk properties, we also compute the formation energy of the $\beta^{\prime}$ alloy. The $\beta^{\prime}$ alloy has a disorder in 10 out of 879 sites with each site having 0.5 chance of being occupied by either Al or Mg~\cite{SampsonPhase}. In our simulations, we consider the two limits where all 10 sites are occupied by either Al or Mg and refer to these as $\beta^{\prime}$(Al) and $\beta^{\prime}$(Mg), respectively. For these two systems, we do not provide KS-DFT results as they are computationally prohibitive. The formation energies for the range of Al-Mg intermetallics are reported in Table~\ref{tab:bulkTrf4}. Our results suggest that the formation energies predicted by RS-OFDFT-FE are in good agreement with PROFESS calculations, and in close agreement with KS-DFT calculations. 
\begin{table*}[htbp]
\caption{\label{tab:bulkTrf4}\small{ Formation energy per atom (eV/atom) of Al-Mg intermetallics calculated using RS-OFDFT-FE, PROFESS, and KS-DFT with TM-NLPS.}}
\begin{ruledtabular}
\begin{tabular}{ccccccc}
 Method & ${\rm Al}_{3}\rm{Mg}$ & ${\rm Mg}_{13}{\rm Al}_{14}$& ${\rm Mg}_{17}{\rm Al}_{12}$
& ${\rm Mg}_{23}{\rm Al}_{30}$ & $\beta^{\prime} ({\rm Al})$ & $\beta^{\prime} ({\rm Mg})$\\ \hline
RS-OFDFT-FE & -0.010 & 0.053 & -0.008 & -0.035 & -0.026 & -0.020\\
PROFESS & -0.011 & 0.052 & -0.011 & -0.034 & -0.029 & -0.023\\
KS-NLPS & -0.007 & 0.061 &-0.027  & -0.019 & - & -\\
\end{tabular}
\end{ruledtabular}
\end{table*}

\subsection{Configurational forces and atomic displacements}~\label{sec:ForceTransferability}
As a next step in our study of the accuracy and transferability of RS-OFDFT-FE, we compute the configurational forces on atoms that are perturbed from their equilibrium positions and compare these with Kohn-Sham DFT calculations. We investigate the accuracy of the forces in both fcc Al and hcp Mg. We begin by considering the relaxed Al fcc unit cell, and the relaxed Mg hcp unit cell. In the relaxed Al fcc unit cell, we perturb the face-centered atom with fractional coordinates $  0 ,\,\,\frac{1}{2}, \,\,\frac{1}{2} $ by 0.1 Bohr in the [0 1 0] direction. In the relaxed Mg hcp unit cell, we perturb the atom with fractional coordinates $\frac{2}{3}, \,\,\frac{1}{3}, \,\,\frac{1}{2}$ by 0.1 Bohr in the [$\bar{2}$ $\bar{1}$ 3 0] direction (directions in hcp Mg are represented using \textit{Miller-Bravais} indices). The configurational forces on the perturbed atoms are computed using RS-OFDFT-FE, and compared against KS-DFT calculations. The computed restoring forces, along [0 $\bar{1}$ 0] for the Al system and along [2 1 $\bar{3}$ 0] for the Mg system, are reported in Table~\ref{tab:tableForceTrf1}. We note that the computed restoring forces from RS-OFDFT-FE are in good agreement with PROFESS and KS-DFT calculations.   

As a more stringent test of accuracy and transferability, we consider the atomic relaxations around a mono-vacancy in fcc Al and hcp Mg. In the case of mono-vacancy in Al, we consider a supercell containing $3 \times 3 \times 3$ fcc Al unit cells and remove an atom to create a mono-vacancy. We calculate the forces on the neighboring atoms of the mono-vacancy, and their relaxation displacements upon ionic relaxation using both RS-OFDFT-FE and KS-DFT calculations. Periodic boundary conditions are employed in these calculations. Table~\ref{tab:tableForceTrf3} reports the computed force and relaxation displacement in Al on the nearest neighboring atom, which experiences the largest ionic force and relaxation. In the case of a mono-vacancy in Mg, we consider a supercell containing $3\times 3\times 2$ hcp unit cells, and Table~\ref{tab:tableForceTrf4} reports the ionic force and relaxation displacement on the neighboring atom that has the largest force in the presence of the vacancy. As is evident from the results, the ionic forces and relaxed displacements for a mono-vacancy in Al and Mg computed using RS-OFDFT-FE are in good agreement with PROFESS, and in close agreement with KS-DFT calculations. These results suggest that the proposed real-space orbital-free DFT formulation provides a good approximation to KS-DFT for Al-Mg materials systems.

\begin{table}[htbp]
\caption{\label{tab:tableForceTrf1}\small{ Restoring force (eV/Bohr) on the perturbed atom in fcc Al and hcp Mg unit cells computed using RS-OFDFT-FE, PROFESS, and KS-DFT calculations.} }
\begin{ruledtabular}
\begin{tabular}{ccccc}
       & RS-OFDFT-FE & PROFESS & KS-BLPS & KS-NLPS  \\
\hline
Al &  0.148 & 0.137 & 0.134 & 0.126  \\ 
Mg &  0.019 & 0.019 & 0.018 & 0.019  \\ 
\end{tabular}
\end{ruledtabular}
\end{table}

\begin{table}[htbp]
\caption{\label{tab:tableForceTrf3} \small{ Ionic forces (eV/Bohr) and relaxation displacement (Bohr) on the nearest neighboring atom to a mono-vacancy in a periodic $3\times 3\times 3$ fcc Al supercell, calculated using RS-OFDFT-FE, PROFESS, and KS-DFT. $f$ and $d$ denote the magnitudes of ionic force and relaxation displacement.  $\angle {\bf f}$  and $\angle {\bf d}$ denote the angles (in degrees) of the force and displacement vectors with respect to the KS-NLPS force and displacement vectors.}}
\begin{ruledtabular}
\begin{tabular}{ccccc}
       & RS-OFDFT-FE & PROFESS & KS-BLPS & KS-NLPS  \\
\hline
$f$  &  0.141 & 0.146 & 0.130 & 0.119  \\ 
$d$  &  9.90\e{-2} & 9.75\e{-2}  & 9.47\e{-2} & 8.90\e{-2}  \\
$\angle {\bf f}$ & 0.00 & 0.00 & 0.00 & 0.00\\
$\angle {\bf d}$ & 0.15 & 0.00 & 0.00 & 0.00\\
\end{tabular}
\end{ruledtabular}
\end{table}
\begin{table}[htbp]
\caption{\label{tab:tableForceTrf4}\small{Ionic forces (eV/Bohr) and relaxation displacement (Bohr) on the nearest neighboring atom to a mono-vacancy in a periodic $3\times 3\times 2$ hcp Mg supercell, calculated using RS-OFDFT-FE, PROFESS, and KS-DFT.}}
\begin{ruledtabular}
\begin{tabular}{ccccc}
       & RS-OFDFT-FE & PROFESS & KS-BLPS & KS-NLPS  \\
\hline
$f$  &  0.059 & 0.060 & 0.053 & 0.046  \\ 
$d$  &  8.26\e{-2} & 8.64\e{-2} & 7.00\e{-2} & 5.83\e{-2}  \\
$\angle {\bf f}$ & 5.11 & 4.73 & 2.75 & 0.0\\
$\angle {\bf d}$ & 5.66 & 5.27 & 3.58 & 0.0\\
\end{tabular}
\end{ruledtabular}
\end{table}

\subsection{Cell-size studies on a mono-vacancy in Al }~\label{sec:Mono-vacancy}
Prior Fourier-space calculations using OF-DFT and WGC Functional~\cite{Ho2007}, and KS-DFT calculations~\cite{Chetty1995} have suggested that cell-sizes containing $\sim 256$ lattice sites are sufficient to obtain a well-converged (to within 3 meV) mono-vacancy formation energy in fcc Al. These Fourier-space calculations, which employ periodic boundary conditions, compute the properties of a periodic array of vacancies. On the other hand, real-space calculations on isolated mono-vacancies in bulk, computed using the recently developed coarse-graining techniques for orbital-free DFT~\cite{Bala2010,QCOFDFT}, suggest that cell-size effects in mono-vacancy calculations are present up to cell-sizes of $\sim 10^3$ atoms. Although both approaches give similar converged vacancy formation energies, this discrepancy in the cell-size effects has thus far remained an open question. 

In order to understand the source of this discrepancy, we conduct a cell-size study of the mono-vacancy formation energy in Al using RS-OFDFT-FE with two types of boundary conditions: (i) periodic boundary conditions on electronic fields; (ii) Dirichlet boundary conditions on electronic fields with values corresponding to that of a perfect crystal. These Dirichlet boundary conditions, which we refer to as bulk Dirichlet boundary conditions, correspond to the scenario where perturbations in the electronic structure due to the mono-vacancy vanish on the boundary of the computational domain, and the electronic structure beyond the computational domain corresponds to that of the bulk. We note that periodic boundary conditions mimic the widely used Fourier-space calculations on point defects, whereas the bulk Dirichlet boundary conditions correspond to simulating an isolated point defect embedded in bulk. We note that the local real-space formulation of orbital-free DFT and the finite-element basis are key to being able to consider these boundary conditions.

We compute the vacancy formation at constant volume as~\cite{Finnis,Gillan1989}
\begin{equation}
E_{vf}= E\left(N-1,1,\frac{N-1}{N}\Omega\right)-\frac{N-1}{N}E\left(N,0,\Omega\right)\,,
\end{equation} 
where $E\left(N,0,\Omega\right)$ denotes the energy of perfect crystal containing $N$ atoms occupying a volume $\Omega$, and $E(N-1,1,\frac{N-1}{N}\Omega)$ denotes energy of a computational cell containing $N-1$ atoms and one vacancy occupying a volume $\frac{N-1}{N}\Omega$. For both periodic boundary conditions and bulk Dirichlet boundary conditions, the lattice site where the vacancy is created is chosen to be the farthest site from the domain boundary.  As we are primarily interested in the cell-size effects of the electronic structure, we do not consider ionic relaxations in this part of our study. Table~\ref{tab:tablemonovac1} shows the unrelaxed mono-vacancy formation energies for different cell sizes computed using RS-OFDFT-FE using both periodic boundary conditions and bulk Dirichlet boundary conditions. We note that the mono-vacancy formation energies using both sets of boundary conditions converge to the same value, and this is also in good agreement with PROFESS and KS-DFT calculations (cf. Table~\ref{tab:tablemonovac2}). However, it is interesting to note that the mono-vacancy formation energies with periodic boundary conditions are well converged (to within 10 meV) by $3\times 3\times 3$ cell-size (108 atoms), whereas we required a $6\times 6\times 6$ cell-size (864 atoms) to achieve a converged formation energy with bulk Dirichlet boundary conditions.

\begin{table}[htbp]
\caption{\label{tab:tablemonovac1}\small{Unrelaxed mono-vacancy formation energies for Al computed using RS-OFDFT-FE with periodic boundary conditions ($E^{p}_{vf}$ in eV) and bulk Dirichlet boundary conditions ($E^{bD}_{vf}$ in eV).}}
\begin{ruledtabular}
\begin{tabular}{cccc}
Cell size & N & $E^{bD}_{vf}$  & $E^{p}_{vf}$ \\
\hline
2x2x2  & 32 & -0.390 & 0.955 \\
3x3x3  & 108 & 0.864 & 0.915 \\
4x4x4  & 256 & 0.971 & 0.908 \\
5x5x5  & 500 & 0.944 &   - \\
6x6x6  & 864 & 0.918 &   - \\
7x7x7  & 1372 & 0.914 &   - \\
\end{tabular}
\end{ruledtabular}
\end{table}

\begin{table}[htbp]
\caption{\label{tab:tablemonovac2}\small{Unrelaxed mono-vacancy formation energies ($E_{vf}$ in eV) for Al computed using PROFESS~\cite{PROFESS2}, and KS-DFT on a $3 \times 3 \times 3$ computational cell. }}
\begin{ruledtabular}
\begin{tabular}{cc}
  & $E_{vf}$ \\
\hline
PROFESS & 0.903\\
KS-DFT-BLPS      & 0.815\\
KS-DFT-NLPS   & 0.811\\ 
\end{tabular}
\end{ruledtabular}
\end{table}

\begin{figure*}[htbp]
\begin{center}
\includegraphics[width=0.9\textwidth]{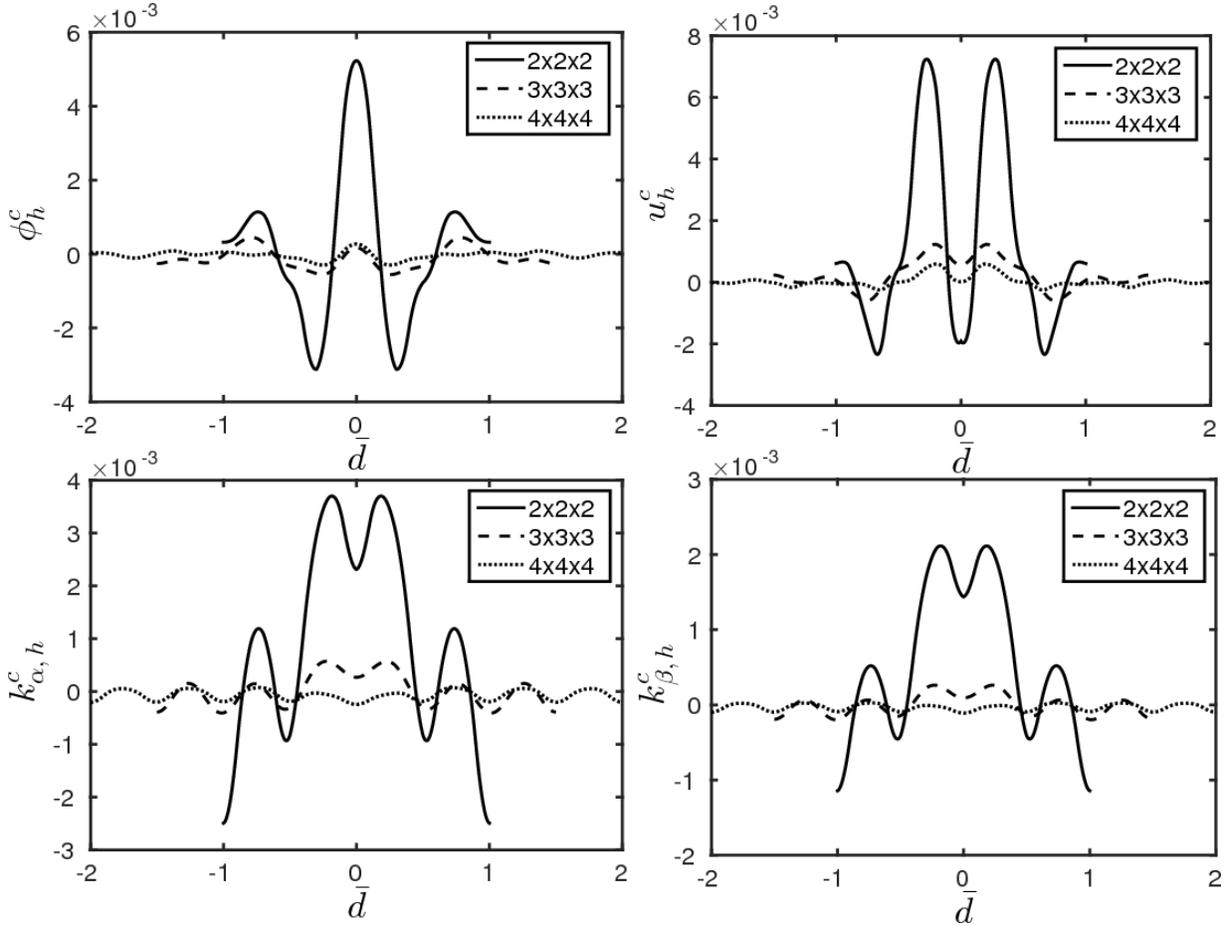}
\caption{\label{fig:monovacCf1}\small{Normalized corrector fields for a mono-vacancy, computed with periodic boundary conditions, along the face diagonal on the computational domain boundary. The abscissa $\bar{d}$ represents a normalized coordinate along the face diagonal. Results for computational cell sizes from $2 \times 2 \times 2$ to $4 \times 4 \times 4$ are shown. }}
\end{center}
\end{figure*}

\begin{figure*}[htbp]
\begin{center}
\includegraphics[width=0.9\textwidth]{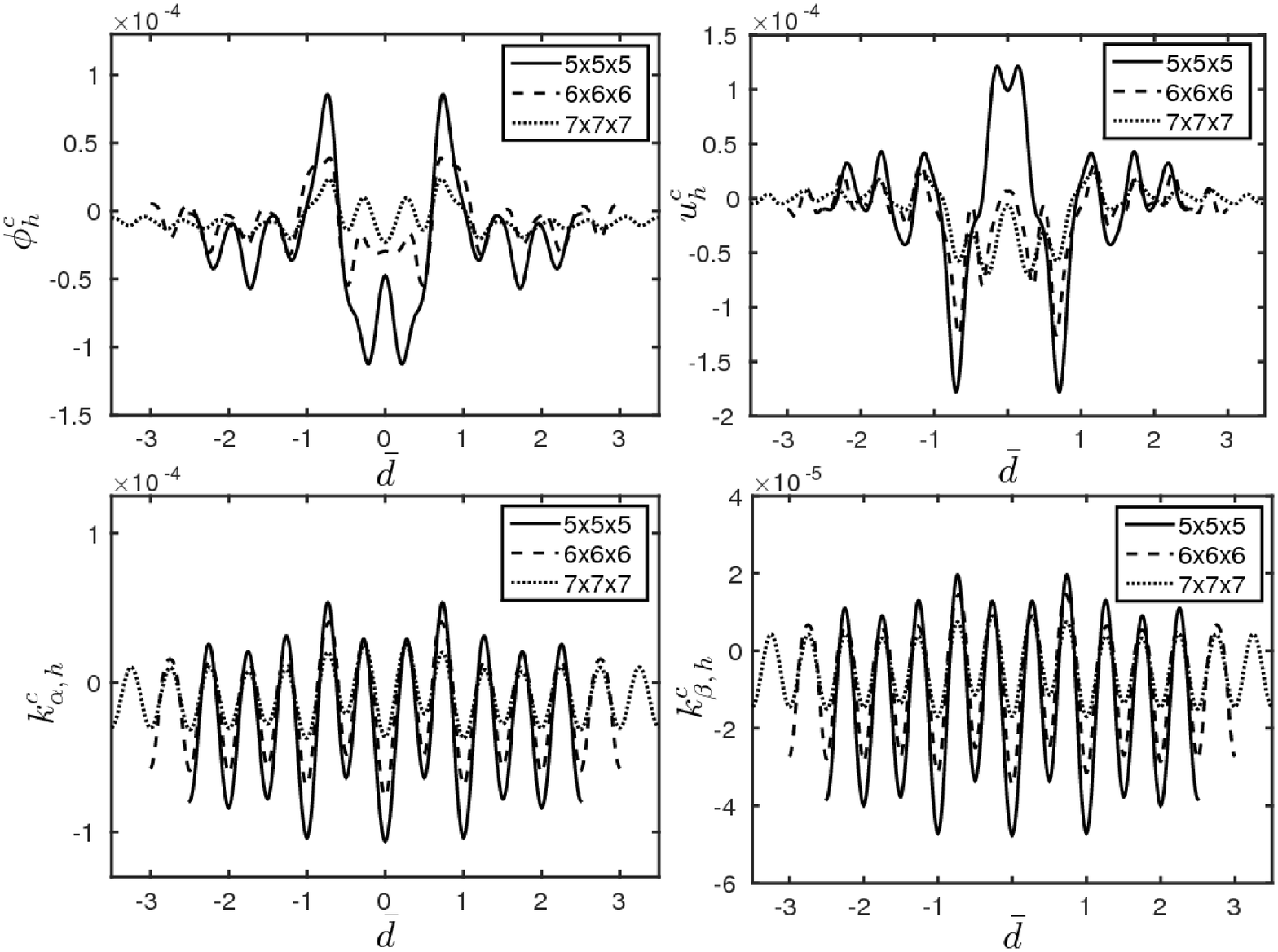}
\caption{\label{fig:monovacCf2}\small{Normalized corrector fields for a mono-vacancy, computed with periodic boundary conditions, along the face diagonal on the computational domain boundary, for cell sizes ranging from $5 \times 5 \times 5$ to $7 \times 7 \times 7$.}}
\end{center}
\end{figure*}

In order to understand this boundary condition dependence of the cell-size effects, we compute the perturbations in the electronic fields due to the presence of the mono-vacancy by subtracting from the electronic fields corresponding to the mono-vacancy the electronic fields of a perfect crystal. To this end, we define the normalized perturbations in the electronic fields computed on the finite-element mesh to be 
\begin{align}\label{eq:correctorFields}
u^{c}_{h}= & \left(u_{h} -u^{p}_{h}\right)/ {\rm v}_{\rm av}\left( u^{p}_{h}\right)\,,\notag \\
\phi^{c}_{h}=& \left(\phi_{h} -\phi^{p}_{h}\right)/{\rm v}_{\rm av}\left( \phi^{p}_{h}\right)\,,\notag\\
k^{c}_{\alpha,h}=& \left(\sum\limits_{j=1}^m\,\omega_{\alpha_j,h}- \sum\limits_{j=1}^m\,\omega^p_{\alpha_j,h}\right)
/{\rm v}_{\rm av}\left( \sum\limits_{j=1}^m\,\omega^p_{\alpha_j,h}\right)\,,\notag\\
k^{c}_{\beta,h}= & \left(\sum\limits_{j=1}^m\,\omega_{\beta_j,h}- \sum\limits_{j=1}^m\,\omega^p_{\beta_j,h}\right)
/{\rm v}_{\rm av}\left( \sum\limits_{j=1}^m\,\omega^p_{\beta_j,h}\right)\,.
\end{align}
In the above, $\{u_{h}, \phi_{h}, \omega_{\alpha_j,h}, \omega_{\beta_j,h} \}$ and $\{u^{p}_{h}, \phi^{p}_{h}, \omega^{p}_{\alpha_j,h}, \omega^{p}_{\beta_j,h} \}$ denote the electronic fields in the computational domain with the vacancy and those without the vacancy (perfect crystal), respectively. ${\rm v}_{\rm av}( . )$ denotes the volume average of an electronic field over the computational cell. As a representative metric, in the definition of $k^{c}_{\alpha,h}$ and $k^{c}_{\beta,h}$ we only consider the kernel potentials corresponding to $K_0$. Figures~\ref{fig:monovacCf1} and ~\ref{fig:monovacCf2} shows the normalized corrector fields for the mono-vacancy, computed using periodic boundary conditions, along the face-diagonal of the periodic boundary. It is interesting to note from these results that the perturbations in the electronic structure due to the vacancy are significant up to $6\times 6\times 6$ computational cells. Thus, although the vacancy formation energy appears converged by $3\times 3\times 3$ computational cell while using periodic boundary conditions, the electronic fields are not converged till a cell-size of $6\times 6\times 6$ computational cell. On the other hand, the cell-size convergence in mono-vacancy formation energy suggested by the bulk Dirichlet boundary conditions is inline with the convergence of electronic fields. These results unambiguously demonstrate that the cell-size effects in the electronic structure of defects are larger than those suggested by a cell-size study of defect formation energies employing periodic boundary conditions. Using bulk Dirichlet boundary conditions for the cell-size study of defect formation energies provides a more accurate estimate of the cell-size effects in the electronic structure of defects, and the extent of electronic structure perturbations due to a defect. Further, while periodic boundary conditions are limited to the study of point defects, bulk Dirichlet boundary conditions can be used to also study defects like isolated dislocations~\cite{Mrinal2015}, whose geometry does not admit periodic boundary conditions. 

\section{Summary}\label{sec:Summary}
We have developed a local real-space formulation of orbital-free DFT with WGC kinetic energy functionals by reformulating the extended interactions in electrostatic and kinetic energy functionals as local variational problems in auxiliary potentials. The proposed real-space formulation readily extends to all-electron orbital-free DFT calculations that are commonly employed in warm dense matter calculations. Building on the proposed real-space formulation we have developed a unified variational framework for computing configurational forces associated with both ionic and cell relaxations. Further, we also proposed a numerically efficient approach for the solution of ground-state orbital-free DFT problem, by recasting the local saddle point problem in  the electronic fields---electron density and auxiliary potential fields---as a fixed point iteration problem and employing a self-consistent iteration procedure. We have employed a finite-element basis for the numerical discretization of the proposed real-space formulation of orbital-free DFT. Our numerical convergence studies indicate that we obtain close to optimal rates of convergence in both ground-state energy and configurational forces with respect to the finite-element discretization. 

We subsequently investigated the accuracy and transferability of the proposed real-space formulation of orbital-free DFT for Al-Mg materials system. To this end, we conducted a wide range of studies on Al, Mg and Al-Mg intermetallics, including computation of bulk properties for these systems, formation energies of Al-Mg intermetallics, and ionic forces in bulk and in the presence of point defects. Our studies indicate that orbital-free DFT and the proposed real-space formulation is in good agreement with Kohn-Sham DFT calculations using both local pseudopotentials as well as non-local pseudpotentials, thus providing an alternate linear-scaling approach for electronic structure studies in Al-Mg materials system. We finally investigated the cell-size effects in the electronic structure of a mono-vacancy in Al, and demonstrated that the cell-size convergence in the vacancy formation energy computed by employing periodic boundary conditions is not commensurate with the convergence of the electronic fields. On the other hand, the true cell-size effects in the electronic structure are revealed by employing the bulk Dirichlet boundary conditions, where the perturbations in the electronic fields due to the defect vanish on the boundary of the computational domain. Our studies indicate that the true cell-size effects are much larger than those suggested by periodic calculations even for simple defects like point defects. We note that the proposed real-space formulation and the finite-element basis are crucial to employing the bulk Dirichlet boundary conditions that are otherwise inaccessible using Fourier based formulations. The proposed formulation, besides being amenable to complex geometries, boundary conditions, and providing excellent scalability on parallel computing platforms, also enables coarse-graining techniques like the quasi-continuum reduction~\cite{QCOFDFT,Mrinal2011} to conduct large-scale electronic structure calculations on the energetics of extended defects in Al-Mg materials system, and is an important direction for future studies.

\begin{acknowledgments}
We gratefully acknowledge the support from the U.S. Department of Energy, Office of Basic Energy Sciences, Division of Materials Science and Engineering under Award No. DE-SC0008637 that funds the Predictive Integrated Structural Materials Science (PRISMS) center at University of Michigan, under the auspices of which this work was performed. V.G. also gratefully acknowledges the hospitality of the Division of Engineering and Applied Sciences at the California Institute of Technology while completing this work. We also acknowledge Advanced Research Computing at University of Michigan for providing the computing resources through the Flux computing platform.
\end{acknowledgments}


\clearpage


\end{document}